\documentstyle[psfig]{mn}
\newif\ifAMStwofonts

\newcommand{\delg}{\delta^{\rm gal.}}
\newcommand{\ste}{\sigma_{{\de}\vert{\te}}}
\newcommand{\sde}{\sigma_{{\te}\vert{\de}}}
\newcommand{\der}{{\rm d}}
\newcommand{\qniec}{\end{document}}
\newcommand{\etal}{{et al.}~}
\newcommand{\de}{\delta}
\newcommand{\te}{\tilde{\theta}}

\newcommand{\lam}{\lambda}
\newcommand{\p}{\partial}
\newcommand{\f}{\frac}
\newcommand{\s}{\sigma}
\newcommand{\bfx}{\bmath{x}}

\newcommand{\bfv}{\bmath{v}}

\newcommand{\calO}{{\mathcal O}}

\newcommand{\calK}{{\mathcal K}}

\newcommand{\eps}{{\epsilon}}
\newcommand{\bc}{\begin{center}}
\newcommand{\be}{\begin{equation}}
\newcommand{\ee}{\end{equation}}
\newcommand{\ba}{\begin{eqnarray}}
\newcommand{\ea}{\end{eqnarray}}
\newcommand{\ec}{\end{center}}
\newcommand{\lan}{\langle}
\newcommand{\ran}{\rangle}

\newcommand{\hmpc}{$h^{-1}\,\mathrm{Mpc}$}
\newcommand{\kms}{{$\mathrm{km}\; \mathrm{s}^{-1}$}}
\newcommand{\cl}{C{\L}97}
\newcommand{\inv}{C{\L}PN}
\title[Density--velocity relation]{Nonlinearity and stochasticity
in the density--velocity relation}

\author[F.~Bernardeau et al.]{F.~Bernardeau,$^1$ 
M.~J.~Chodorowski,$^2$ 
E.~L.~{\L}okas,$^2$ 
R. Stompor$\,^{3,2}$ \cr
and A.~Kudlicki$\,^2$\\ 
$^1$Service de Physique Th\'{e}orique, CE de
Saclay, F-91191 Gif-sur-Yvette, Cedex France \\
$^2$Copernicus Astronomical Center, Bartycka 18,
00-716 Warsaw, Poland\\ 
$^3$Center for Particle Astrophysics, University of California,
Berkeley, California, USA}
\begin{document}
\maketitle
\begin{abstract}

We present results of the investigations of the statistical properties
of a joint density and velocity divergence probability distribution
function (PDF) in the mildly non-linear regime. For that purpose we
use both perturbation theory results, extended here for a top-hat
filter, and numerical simulations.

In particular we derive the quantitative (complete as possible up to
third order terms) and qualitative predictions for constrained
averages and constrained dispersions -- which describe the
nonlinearities and the stochasticity properties beyond the linear
regime -- and compare them against numerical simulations. We find
overall a good agreement for constrained averages; however, the
agreement for constrained dispersions is only qualitative. Scaling
relations for the $\Omega$-dependence of these quantities are
satisfactory reproduced.

Guided by our analytical and numerical results, we finally construct a
robust phenomenological description of the joint PDF in a closed
analytic form. The good agreement of our formula with results of
N-body simulations for a number of cosmological parameters provides a
sound validation of the presented approach.

Our results provide a basis for a potentially powerful tool with which
it is possible to analyze galaxy survey data in order to test the
gravitational instability paradigm beyond the linear regime and put
useful constraints on cosmological parameters. In particular we show 
how the nonlinearity in the density--velocity relation can be used to
break the so-called $\Omega$-bias degeneracy in cosmic
density--velocity comparisons.

\end{abstract}

\section{Introduction}
\label{sec:intro}

Comparisons between cosmic density and peculiar velocity fields of
galaxies have become already a small industry in cosmology (e.g.,
Strauss \& Davis 1988; Yahil 1988; Kaiser \etal 1991; Dekel \etal
1993; Hudson 1993; Hudson \etal 1994; Davis, Nusser \& Willick 1996;
Riess \etal 1997; Willick \etal 1997; da Costa \etal 1998; Sigad \etal
1998; Willick \& Strauss 1998). The goal {of this pursuit} is to test
the gravitational instability hypothesis for the formation of
structure in the Universe and to measure the cosmological parameter
$\Omega$. In the paradigm of gravitational instability the density and
the velocity fields are tightly related and the relation between them
depends on $\Omega$. In the linear regime, i.e., when the density
fluctuations are significantly smaller than unity, this relation
{reads,} \be \de(\bfx) = - f(\Omega,\Lambda)^{-1} \nabla \cdot \bfv(\bfx)
\,. \label{eq:i1} \ee Here, $\de$ is the mass density fluctuation
field, $\bfv$ is the velocity field, $f(\Omega,\Lambda) \simeq \Omega^{0.6}$
and we express distances in units of \kms.  However, the {derived
amplitude of the} density fluctuations from current redshift surveys
(e.g., Fisher \etal 1995) and from the {\sc potent} reconstruction of
density fields (Dekel \etal 1990 and Bertschinger \etal 1990), {goes
somewhat beyond} the linear regime. For example, the density contrast
in regions like the Great Attractor or Perseus-Pisces is about unity
even when smoothed over scales of $1200$ \kms\ (Sigad \etal
1998). Future redshift surveys and peculiar velocity catalogs are
expected to provide reliable estimates of density and velocity fields
on scales where nonlinear effects are certainly non-negligible {and
may lead to interesting consequences such as breaking the degeneracy
between $\Omega$ and bias (Chodorowski \& {\L}okas 1997, hereafter
\cl; this paper; Chodorowski, in preparation) and therefore need to be
accounted for in performed analyses.}  To date, there have been
several attempts to construct a mildly nonlinear extension of
relation~(\ref{eq:i1}). Those were either based on various analytical
approximations to nonlinear dynamics (Nusser \etal 1991; Bernardeau
1992b; Gramann 1993; Mancinelli \& Yahil 1995; Chodorowski 1997), or
N-body simulations (Mancinelli \etal 1994; Ganon et al., in
preparation).

The aim of this paper is to describe {\em quantitatively\/} the the
density versus velocity-divergence relation (DVDR) at mildly nonlinear
scales.\footnote{We define a density field to be mildly nonlinear
if the r.m.s. density fluctuation is a significant fraction of, but
still smaller than, unity. Then the mildly nonlinear scales in the
Universe are these about or greater than 8
\hmpc.} We focus on that regime because it is explored in current
analyses of the observational data, which are commonly smoothed over
scales chosen to ensure that the r.m.s.\ density fluctuation, though
significant, is below unity. Also this is the domain of applicability
of PT, permitting us applying some of the recent results obtained
within its framework and relevant to our goal.

Recently, \cl\ and Chodorowski \etal (1998, hereafter \inv) derived
the DVDR up to third order in perturbation theory (PT), assuming
Gaussian initial conditions. Specifically, \cl\ derived the `forward'
relation (the density in terms of the velocity divergence), while
\inv\ derived the `inverse' relation (the velocity divergence in terms
of the density) and the scatter in both relations. Due to the scatter,
the inverse relation is not identical with a mathematical inversion of
the forward one. The coefficients in these relations were calculated
for the case of Gaussian smoothing of the evolved fields which is also
applied in the analysis of observational data.  For large smoothing
scales, N-body results are expected to converge to the results of \cl\
and \inv.  Indeed, the predicted values of the coefficients proved to
be in qualitative agreement with the results of Ganon et al. (in
preparation), who tested similar functional forms of the relations
against N-body simulations. On the other hand, for small smoothing
scales (at which the r.m.s. density fluctuation is close to unity),
contributions from orders higher than third may become significant. 

In this paper, instead of developing a PT methodology beyond the third
order, we make use of N-body simulations. Yet, in the mildly nonlinear
regime, our approach is still superior to those based solely on means
of N-body simulations.  The asymptotically exact formula, we wish to
modify here is the complete third-order relation of \cl\ and \inv, not
the linear expression given by Equation~(\ref{eq:i1}). Indeed, as we
will see the modification to the third-order results concerning the
mean (the `forward' and `inverse') relations is quite modest. In other
words, the third order mean relations prove to be already good
approximations. However, since the mildly nonlinear DVDR is not
deterministic, its full description should include not only the
constrained averages but the whole joint Probability Distribution
Function (PDF) for density and velocity divergence. We construct here
this PDF in a closed analytic form.

In order to establish accurately the DVDR from N-body simulations, one
has to treat properly the final velocity field, determined in a
simulation only at a set of discrete points (final
positions). Commonly used smoothing procedures lead to effectively a
mass averaged velocity field, which induces spurious velocity
gradients. The problem of proper volume-weighting of the velocity
field was solved for a top-hat smoothing by Bernardeau \& van de
Weygaert (1996) using Voronoi and Delaunay tessellations. Therefore,
here we will investigate the DVDR for top-hat smoothed fields. The
results of \cl\ and \inv\ were obtained for Gaussian smoothing;
consequently we extend them for top-hat smoothing. The algorithm of
Bernardeau \& van de Weygaert (1996) can be possibly modified so as to
apply it to Gaussian-smoothed fields. This will be addressed elsewhere
(Chodorowski \&\ Stompor, in preparation).

In sum, new parts of this investigation are:
\begin{enumerate}
	\item Calculation up to third order in PT of the coefficients
	of the mean relations and of the scatter in both, for top-hat
	smoothing;
	\item Comparison of PT predictions to N-body simulations in
	the mildly nonlinear regime;
	\item Construction of the joint PDF for density and velocity
	divergence.
\end{enumerate}
Also, we will show how the nonlinearity in the DVDR can be used to
break the so-called $\Omega$-bias degeneracy in comparisons between
large-scale density fields of galaxies and the fields of their
peculiar velocities.

In Sect. \ref{PTresults}, we present the results obtained
from PT for the constrained averages. Sect. \ref{NBresults} is
devoted to the presentation of numerical results. We first confront
them {against PT-derived predictions and subsequently guided by both
kind of results} construct a phenomenological 
description of the joint density--velocity PDF.
The technical 
calculations are deferred to appendices. These contain in particular the
explicit expressions of the coefficients that intervene in PT.

\section{Perturbative results}
\label{PTresults}
\begin{figure}
  \begin{center}
    \psfig{figure=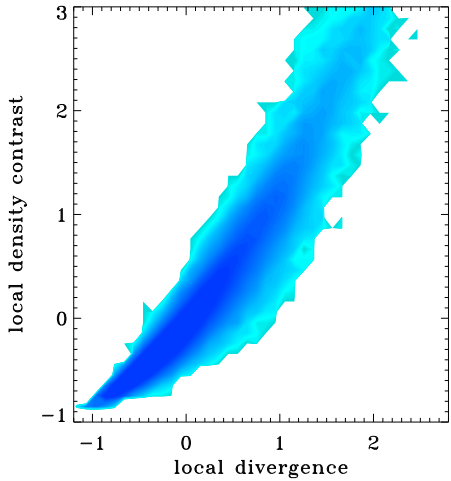,height=7.cm} 
  \end{center}
\caption{A typical joint PDF obtained in numerical simulations
(see text). The smoothing scale is 15$h^{-1}$Mpc, and the contour
plot is made in logarithmic scale.}
\label{fig:pdf}
\end{figure}

\subsection{General features}
At large scale, it is expected from linear theory that the local
density contrast and the velocity divergence are proportional
(Eq. \ref{eq:i1}).  However, as soon as nonlinear effects come into
play, and this is expected to happen when the variance reaches a
fraction of unity, this relation is expected to fail. {The
true relation departs from that given by Equation~(\ref{eq:i1}) in a number
of ways. First, its linearity evidently does not hold any more (see
Fig.~\ref{fig:pdf}). Secondly, and more importantly, no more a
one-to-one correspondence exists between local density and local
velocity divergence.} Non-local effects introduce a scatter in
this relation.  As a result, the description of the relation
between the local density and the local convergence should be made
with the construction of the joint Probability Distribution Function
(PDF) of these two quantities.  Let us define the velocity divergence
field,

\be
\theta(\bfx) = {\nabla \cdot \bfv(\bfx)}
\,, \label{eq:the} 
\ee 
and the {\em scaled\/} velocity divergence with
\be
\te(\bfx) = -f(\Omega,\Lambda)^{-1}\ \theta 
\,. \label{eq:theta} 
\ee
Note the minus sign and the factor $f(\Omega,\lambda)^{-1}$ multiplying the
velocity divergence. Linear relation~(\ref{eq:i1}) is then simply 
$\de=\te$. 
In the linear regime this PDF, $P_{\rm joint},$ therefore reads,
\be
P_{\rm joint}^{\rm linear}(\de,\te)\ \der\de\ \der\te
=\exp\left[{-{\te^2\over2\,\sigma^2}}\right]\ 
\delta_{\rm Dirac}(\de-\te){\der\de\ \der\te\over\sqrt{2\,\pi\,\sigma^2}},
\label{Plin}
\ee 
{where $\s$ is the r.m.s.\ fluctuation of the density field.}
Non-linear couplings will not only introduce some non-Gaussian
features in the PDF of {each of the variables,} they will also
change the Dirac delta function ($\delta_{\rm Dirac}$) in a more
complicated function of a finite width. {As long as the departure
from the linear regime is small, generic features can be inferred from
Perturbation Theory (Bernardeau 1992b). The r.m.s.\ value of the
scatter around the mean relation between $\de$ and $\te$ is expected
to be of the order of $\sigma^2$. (As stressed by Chodorowski 1997,
this can be interpreted as an effect due to the couplings of the local
density with the shear field.) Therefore, for $\sigma$ beyond
linear regime but still below unity the most probable values of
$(\de,\te)$ form an elongated region in a $(\de,\te)$ plane as it is
evident in Fig. \ref{fig:pdf}.}

The mean trend of the DVDR is described by constrained averages,
either by the mean density given $\te$ or, equivalently, the mean
$\te$ given $\de$.  The dispersion around that mean trend provides us
with the complementary information that is the amplitude of the
scatter.  All these features are a priori accessible to analytic
calculations in PT; some of them have been done in the previous papers
{(Bernardeau 1992b, \cl, \inv)}. The results however are
dependent on the chosen window function. Since filtering in Voronoi
tessellations has only been implemented for top-hat window function we
decide to focus our presentation on this case.

\subsection{Mean relations}
The generic formal expression of the mean relation is a standard
calculation in statistics. We sketch its derivation in Appendix
\ref{app:pdf}.  In general the constrained mean density, $\lan \de
\ran_{\vert_{\te}}$, is a function of both $\te$ and $\sigma$.  These
two variables are a priori small and of the same order, however to
allow us the description of rare events it is convenient to separate
the expansion in $\te$ from the one in $\sigma$.  An important result
obtained by Bernardeau (1992b) is the expression of the constrained
density as a function of $\te$ in a case of a vanishing variance. The
constrained means have been found to be given by the $\de$--$\te$
relation exhibited in the spherical collapse model. A simplified
useful expression (which is strictly valid in the limit $\Omega \to
0$) is given by,

\be
\lan \de \ran_{\vert_{\te}}=
\left(1+{2\te\over 3}\right)^{3/2}-1,\ \ \ \ {\rm for}\ \ \sigma\to 0.
\label{B92}
\ee

{Though this result was obtained without taking into account the
filtering effect, progress made in understanding of 
filtering has shown that in a case of a top-hat filter the
filtering effects can be described as a simple mapping from the
Lagrangian to the Eulerian space (Bernardeau 1994b). As a consequence
{\em the relation (\ref{B92}) is expected to be valid for
hop-hat filtering as well.}
}  

When the limit $\sigma\to 0$
is dropped only a finite number of terms in a joint expansion in
$\sigma$ and $\te$ is known.  They have been obtained in several
recent papers {(Bernardeau 1992b, \cl, \inv)}.
Up to the third order the mean $\de$ given $\te$ 
can be written generically as,
\be 
\lan \de \ran_{\vert_{\te}} = a_1(\sigma_{\te})\ \te + 
a_2 (\te^2 - \s_{\te}^2) + a_3 \te^3,
\label{eq:for}
\ee
and mean $\te$ given $\de$,
\be
\lan \te \ran_{\vert_{\de}} = r_1(\sigma_{\de})\ \de + r_2 (\de^2 - \s_{\de}^2) +
r_3 \de^3 
\,. \label{eq:inv}
\ee
Here, $\s_{\te}^2$ and $\s_{\de}^2$ are the variances of the density and
the scaled velocity divergence field respectively and these
$\sigma^2$ terms are included so that the global
averages of $\de$ and $\te$ vanish. To have a consistent expansion
up to the third order both $a_1$ and $r_1$ have to be calculated up to
order $\sigma^2$ whereas the other coefficients can be
computed at their leading order. The coefficient $a_2$ was computed by
Bernardeau (1992b) for a top-hat filter
and it is easy to check that $r_2=-a_2$. More generally 
the coefficients $a_m$ and $r_m$ were derived by \cl\ and \inv\
respectively, and their values were explicitly calculated for a
Gaussian window function. Up to the third order, 
the coefficients $a_2$, $a_3$, $r_2$ and
$r_3$ do not depend on the normalization of the power spectrum but
they depend slightly on the smoothing scale through the power
spectrum index (unless it is a power law).
In Appendix~\ref{app:coeff} we explicitly
compute the values of the coefficients for a top-hat window
function. Unlike for a Gaussian smoothing, their values turn out
additionally not to depend on the spectral index, as expected
if the picture of the Lagrangian to Eulerian mapping is correct. 
Specifically we have,
\be
a_2 = \frac{4}{21} \simeq 0.190 \,, \qquad 
a_3 =  - \frac{40}{3969} \simeq -0.0101 
\label{eq:a2a3}
\ee
and
\be
r_2 = - \frac{4}{21} \simeq - 0.190 \,, \qquad 
r_3 = \frac{328}{3969} \simeq 0.0826 \,.
\label{eq:r2r3}
\ee
It is worth noting that $a_2$ and $a_3$ are very close 
to those obtained 
{Taylor expanding eqn. (\ref{B92}) in terms of $\te$}, indicating extremely weak
$\Omega$-dependence of the relation between density and scaled
velocity. 
The values of the coefficients $a_1$ and $r_1$ are at the leading
order both equal to unity. Their next-to-leading-order corrections, as
predicted by PT, are proportional to the variance of the respective fields
[equations~(\ref{a1}) and~(\ref{r1})]. In fact, up to the third
order it is sufficient to consider the linear variance of the fields,
which is identical, and will be denoted $\s^2$. The coefficients of
proportionality of the corrective terms depend on the spectral
index, similarly to the case of a Gaussian smoothing. We list the
values of $(a_1-1)/\s^2$ and $(r_1-1)/\s^2$ as functions of the
spectral index in Table~\ref{taba1}.

\subsection{Scatter in the relations}

The mean relations, both forward and inverse, have a scatter. The
r.m.s.\ value of the scatter is equal to the square root of the
conditional variance, 
\ba
\ste \equiv \lan (\de - \lan \de \ran_{\vert_{\te}})^2
\ran_{\vert_{\te}}^{1/2} \ \ \ {\rm (forward)},  \nonumber\\
\sde \equiv \lan (\te - \lan \te \ran_{\vert_{\de}})^2
\ran_{\vert_{\de}}^{1/2} \ \ \ {\rm (inverse).} \nonumber
\ea

Hence up to the next-to-leading order (\inv), 
\be
\ste^2 = b_0 \s_{\te}^4 + b_1 \s_{\te}^4 \te + b_2 \s_{\te}^2 \te^2 + 
\calO(\s_{\te}^6) 
\label{stet}
\ee
and 
\be
\sde^2 = s_0 \s_{\de}^4 + s_1 \s_{\de}^4 \de + s_2 \s_{\de}^2 \de^2 + 
\calO(\s_{\de}^6)\,.
\label{sdet}
\ee
PT predicts furthermore that
\be
s_0 = b_0 \qquad \hbox{and} \qquad s_2 = b_2\;.
\label{b0res}
\ee 
Explicit formulas for the coefficients $b_0$ and $b_2$ were derived
by \inv. 
In expression~(\ref{stet}), the terms $b_0 \s_{\te}^4$ and $b_2 \s_{\te}^2
\te^2$ are of the leading, $\calO(\s_{\te}^4),$ order (as $\te \sim
\s_{\te}$), while the term $b_1 \s_{\te}^4 \te$ is of the next-to-leading
order, $\calO(\s_{\te}^5)$. We have included this term because for a
top-hat filter the value of the coefficient $b_2$ is exactly
zero\footnote{In the case of a Gaussian smoothing, $b_2$ is not
exactly zero but still very small.} (see Appendix~\ref{app:coeff}).
Therefore, the lowest-order non-zero contribution to the
$\te$-dependence of the constrained variance is already provided by a
next-to-leading-order term. In Appendix~\ref{app:coeff} we calculate
the values of the coefficient $b_0$ for power-law spectra. The results
are presented in the last column of Table~\ref{taba1}. 
We found that the expression for $b_1$ (and $s_1$) contains
perturbative contributions up to the third-order.
However, the mixed moments which enter the calculations (with terms 
of the form $\langle v_1 v_2 v_3\rangle$ where $v$ stands
for either $\de$ or $\te$)
are of extra complexity in comparison to any other moments
computed here. In the present paper we will not therefore try to
predict the value of $b_1$ or $s_1$. 

\section{Numerical results}
\label{NBresults}

The aim of this section is twofold. We want
\begin{itemize}
\item to check, qualitatively 
and quantitatively, the PT results
presented in the previous Section. 
\item to build a complete phenomenological description of the joint PDF
of the local density and velocity divergence that would be valid for
typical events, $\de$ of order $\s_{\de}$ and
$\te$ of order $\s_{\te}$, as well as for the rare event tails.
\end{itemize}

Except for the formula~(\ref{B92}), all
perturbative results were derived
for $\de\sim\s_{\de}$ and $\te\sim\s_{\te}$. 
For applications of these results to
current density--velocity comparisons this assumption is
justified. The volume of current peculiar velocity surveys is too
small, sampling too sparse and noise too big to detect unambiguously
`rare events' in the observed velocity field. 
However, in current
N-body numerical simulations the simulation volume can be made
significantly greater, sampling is dense and there is essentially no
noise in the velocity field traced by particles. Therefore, the use
N-body simulations enables one to leave the regime of `typical events'
in the $(\de,\te)$ plane, i.e., to study `rare events' as well. This
is not only of academic value, since with accumulation of new velocity
data and discovery of new distance indicators we may hope the observed
velocity field to improve qualitatively in the future. 

Here we use a set of numerical N-body simulations to study the DVDR in
the mildly nonlinear regime (i.e., $\s_{\de} < 1$), but including also
its behaviour in the low- and high-density tail. This will be
especially important for constructing the joint PDF for the density
and the velocity divergence. A brief description of the N-body
simulations as well as numerical algorithms employed in their analysis
are given in Subsection~\ref{sub:nbody}. In Subsection~\ref{sub:PTver}
and \ref{sub:pheno} we study the constrained averages and the
constrained dispersions, respectively, and compare them to the PT
predictions. Subsection~\ref{sub:PDF} provides details of how we build
a phenomenological description of the joint PDF using both PT and
simulation derived information. 


\subsection{The simulations and the numerical algorithms}
\label{sub:nbody}

The N-body simulations were carried out using a gravity solver based
on AP$^3$M code developed by Couchman (Couchman 1991), which we
adapted for use in cases with arbitrary negative curvature (open
models) and/or cosmological constant (see for example Peebles 1980 for
general formulas and Peacock \&\ Dodds 1996 for a description of
necessary changes).

In each case, we used $128^3$ particles within the cubic
$200h^{-1}$Mpc size box with a $128\times128\times128$ FFT (Fast
Fourier Transform) grid.  Both amplitudes and phases of an initial
density field were randomly drawn from the Rayleigh and flat
distribution respectively and initial positions and velocities of the
particles were fixed using standard Zel'dovich movers accordingly.
The starting time of the simulations was chosen to ensure that an
average initial particle displacement was less than one third of the
elemental  cell size with the power on that scale being initially of
the order of $2\div5\times10^{-2}$. The gravitational force smoothing
was always set equal to the single cell size at the beginning of the
simulation, kept subsequently constant in physical units to become
eventually frozen on one tenth of the comoving cell size ever since
reaching that limit (Couchman 1991).

The equations of motion were then integrated in 500 time-steps over an
expansion factor of 30 and ended at the time when a (ensemble
averaged) standard deviation of linear mass perturbations reached a
required value at a given scale (usually unity on the scale of $8h^{-1}$Mpc).

We had at our disposal 5 simulations done for power law spectra
of spectral index $n=-1.5$ and corresponding to different cosmological
parameters, either for an Einstein-de Sitter Universe, or with
$\Omega=0.3$ and $\lambda=0$, or with $\Omega=0.3$ and $\lambda=0.7$ at
the final stage of the simulation. We choose index $n=-1.5$ so that
the results of PT that involve loop terms are finite (e.g. Table \ref{taba1}).
See discussions in the following.

In each simulation we applied a top-hat filter of different radius.
In each case the filtered quantities were obtained on
$50^3$ grid points. The local filtered velocity divergence
was obtained using the Voronoi tessellation technique developed
by Bernardeau \& van de Weygaert (1996). We have checked that the
number of tracers we have used (about 60,000) was sufficient
to provide us with robust results for the scales of interest.

\subsection{The constrained averages}
\label{sub:PTver}

Fig. \ref{fig:pdf} shows a typical result for the joint PDF.
{}From such data it is easy to compute the constrained averages. 
We then use the formulas (\ref{eq:for}) and (\ref{eq:inv})
as parametric functions leaving the three coefficients and
the variance as free parameters to fit the 
measured expectation values. The fitting procedure
uses a weighting function centered on either $\de=0$ or $\te=0$
with a width of the order of $\s$. We have checked that our
fitting method provided us with robust results for the coefficients.
To illustrate the effect of cosmic variance on the values of the
estimated parameters, instead of plotting their values averaged over 5
simulations, we decided to make combined plots of the coefficients
derived from each simulation. 

\begin{figure}
  \begin{center}
    \psfig{figure=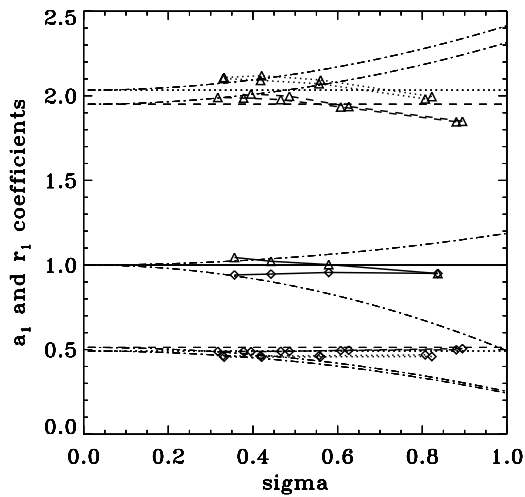,width=7.5cm}
    \end{center}
\caption{
The best-fit values of the coefficients $a_1$ (triangles) and $r_1$
(diamonds) as determined for a set of different simulations
(interpolated by thick solid lines for Einstein-de Sitter, 
dotted lines for open Universes with $\lambda=0$, dashed lines for non-zero
$\lambda$ Universes). In this case and for this plot only the displayed
coefficients are those describing the relation between 
$\de$ and the (minus) true divergence, similarly to 
Eqs. (\ref{eq:for}, \ref{eq:inv}), to make their
$\Omega$ and $\Lambda$ dependence apparent. Note a very weak
dependence on $\Lambda$, as expected. The thin horizontal lines
are the linear theory predictions. The thin dot-dashed
lines are the PT predictions at the one loop order (see Table C1).
}
\label{fig:linCoef}
\end{figure}

\begin{figure}
  \begin{center}
    \psfig{figure=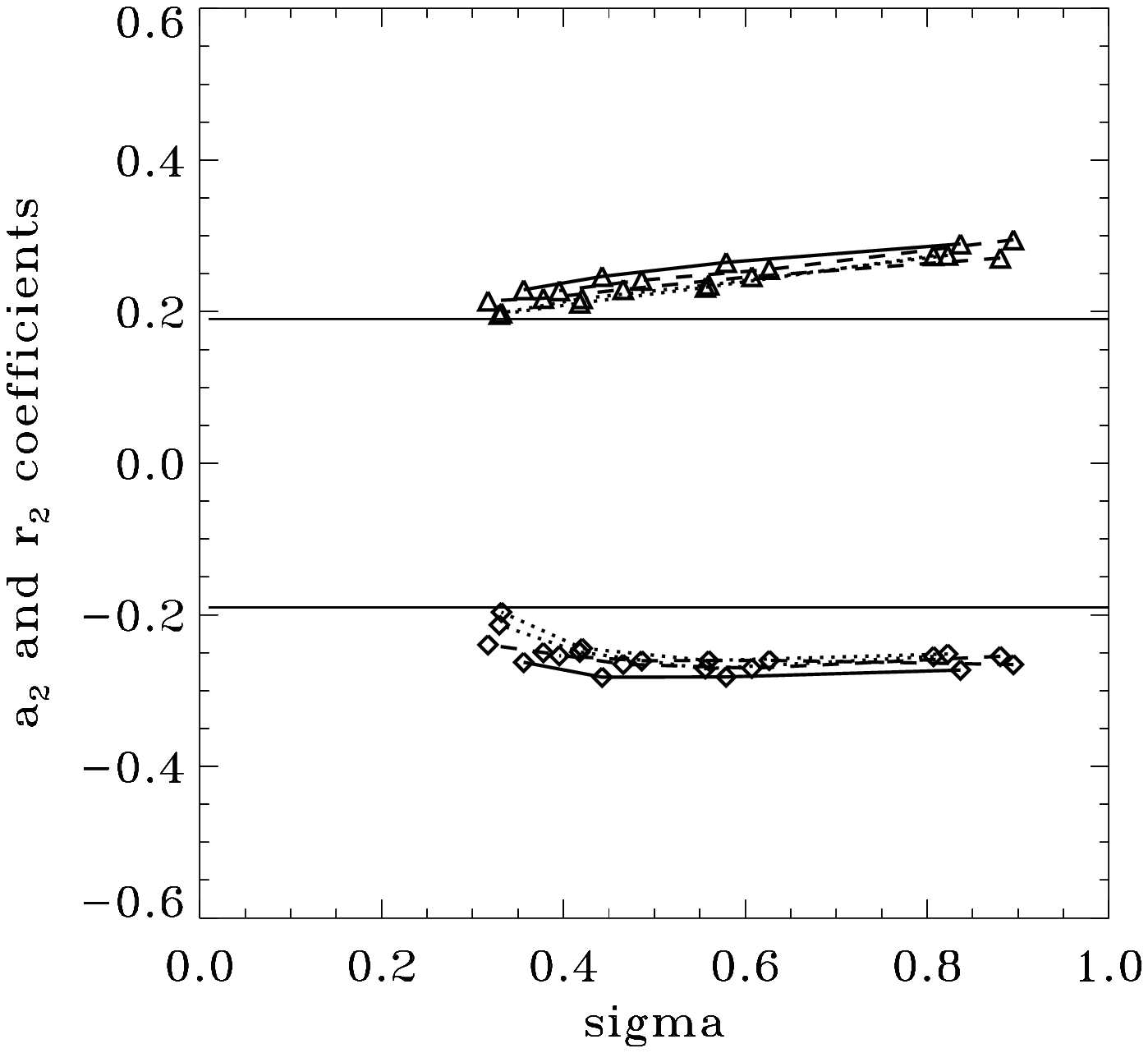,width=6.5cm} 
    \psfig{figure=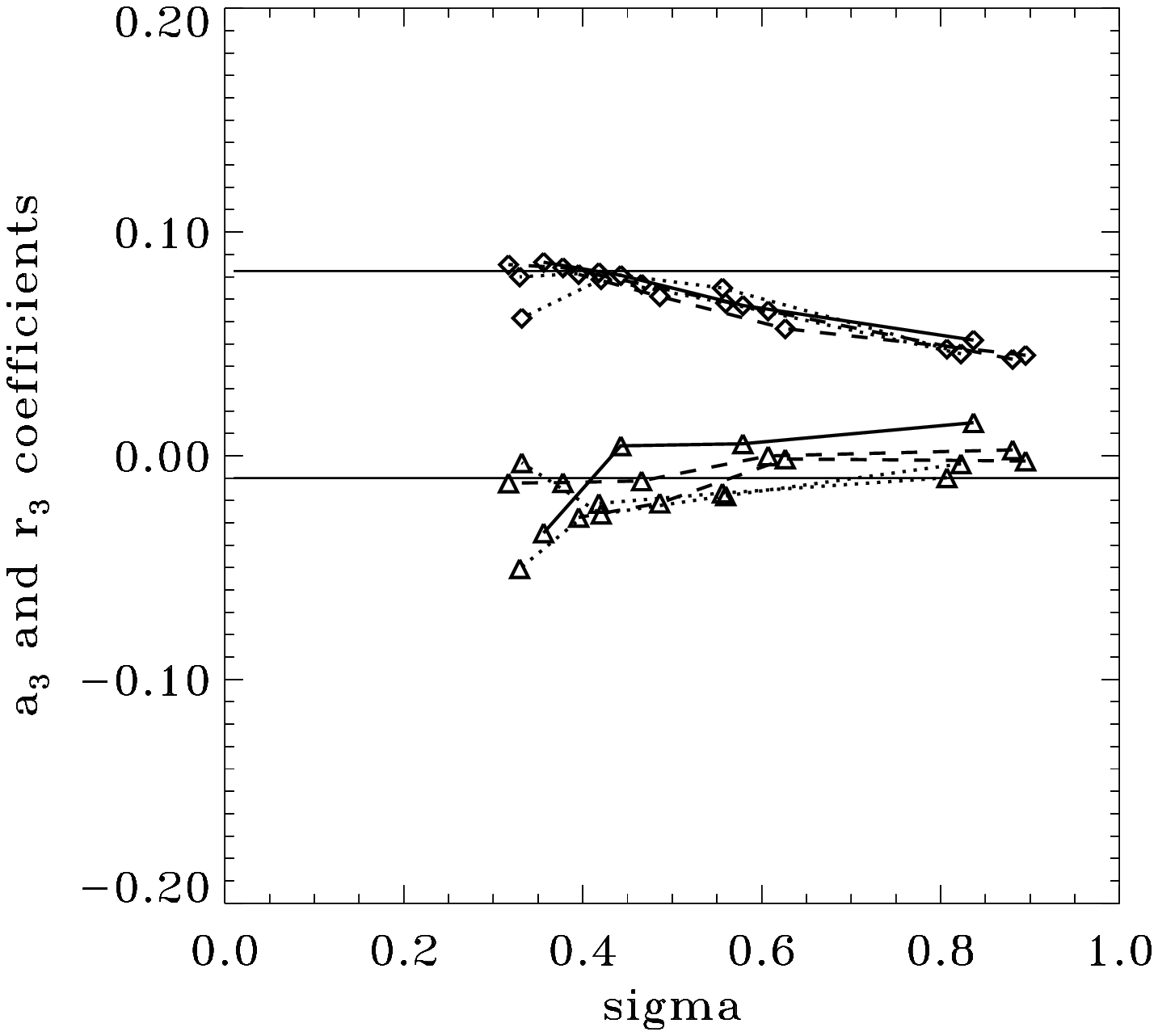,width=6.5cm} 
  \end{center}
\caption{{The best-fit values of the coefficients $a_2$ and $r_2$ (top panel)
and $a_3$ and $r_3$ (bottom panel) as
determined for different simulations (conventions are as in
Fig. 2). The presented here coefficients are those 
at the quadratic and cubic terms
in the $\de$--$\te$ relation.}
}
\label{fig:arCoef}
\end{figure}

In Fig. \ref{fig:linCoef} we present the resulting values of the
linear coefficient between the local density contrast and the minus
divergence (in this case, not the scaled divergence). It explicitly
exhibits the $f(\Omega,\Lambda)$ dependence that we expect for different
cosmologies.  Fig. \ref{fig:arCoef} shows the fitted values of
$a_2$, $r_2$, $a_3$ and $r_3$ for different cosmologies and smoothing
scales. In this case we present the results for the fitting
coefficients between $\de$ and the scaled divergence $\te$.  Note that
with these variables we expect the coefficients $a_n$ and $r_n$ to be
independent of a cosmological model. This property is indeed observed
in the results. Solid horizontal lines in Figs.~\ref{fig:linCoef}
and~\ref{fig:arCoef} represent the PT prediction for the coefficients
calculated at leading order. We can see that for small values of the
variance the coefficients derived from the simulations converge to
their leading-order PT values. For large values of the variance, the
coefficients slightly, but systematically deviate from these values. 

The next-to-leading-order corrections to the coefficients $a_k$ (or
$r_k$) are given by so-called one-loop terms involving perturbative
contributions of $(k+2)$-th order. Therefore, here we were able to
compute a one-loop correction to the linear value of the coefficients
$a_1$ and $r_1$ only (since we performed a complete 
third-order calculation.). {}
The results are plotted as thin dot-dashed lines in
Fig.~\ref{fig:linCoef}, and are rather puzzling. {}
While PT predicts a significant departure of $a_1$ and $r_1$ 
from their linear values, the
coefficients derived from simulations remain in fact essentially
constant as a function of $\sigma$. It does not seem to be a numerical
problem. This failure of next-to-leading order predictions
is possibly due to the existence of divergences in the PT
calculation.
This is a difficulty
that has been encountered in other one-loop calculations in PT (see
Jain \& Bertschinger 1994,
Scoccimarro \& Frieman 1996, and this paper).\footnote{However,
for some of us it still remains a puzzle, since we have at least a
partial physical explanation for the presence of the term $\propto
\s_{\te}^2 \te$ in the forward relation. In the Zel'dovich
approximation, the Eulerian density is a third-order form in
(first-order) velocity derivatives, and one of the third order terms
can be cast to the form $\propto \Sigma^2 \te$, where $\Sigma^2$ is
the shear scalar (Chodorowski 1998). One of the contributions to
 $\lan \de \ran_{\vert_{\te}}$ is then proportional to the constrained
average $\lan \Sigma^2 \te \ran_{\vert_{\te}}$, which, since at
leading order the shear is independent from the divergence, is $(2/3)
\s_{\te}^2 \te$ !}

The cosmic variance is generally quite small.
Not surprisingly, the effect of the
cosmic variance is the strongest for the coefficients $a_3$ and $r_3$,
and on the largest scales (the smallest $\sigma$). 

\subsection{The constrained dispersions}
\label{sub:pheno}

\begin{figure}
  \begin{center}
    \psfig{figure=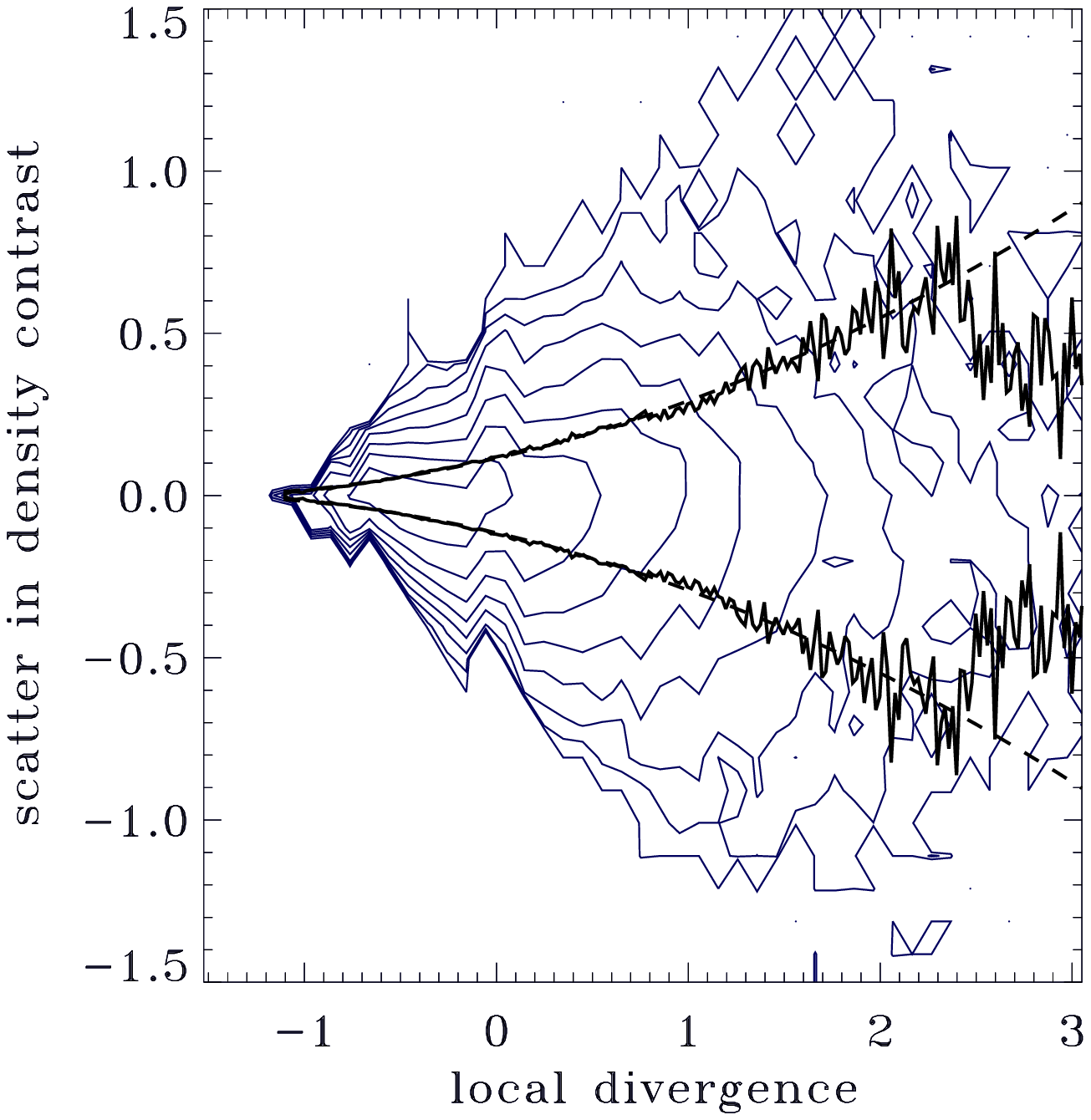,width=7.cm} 
    \psfig{figure=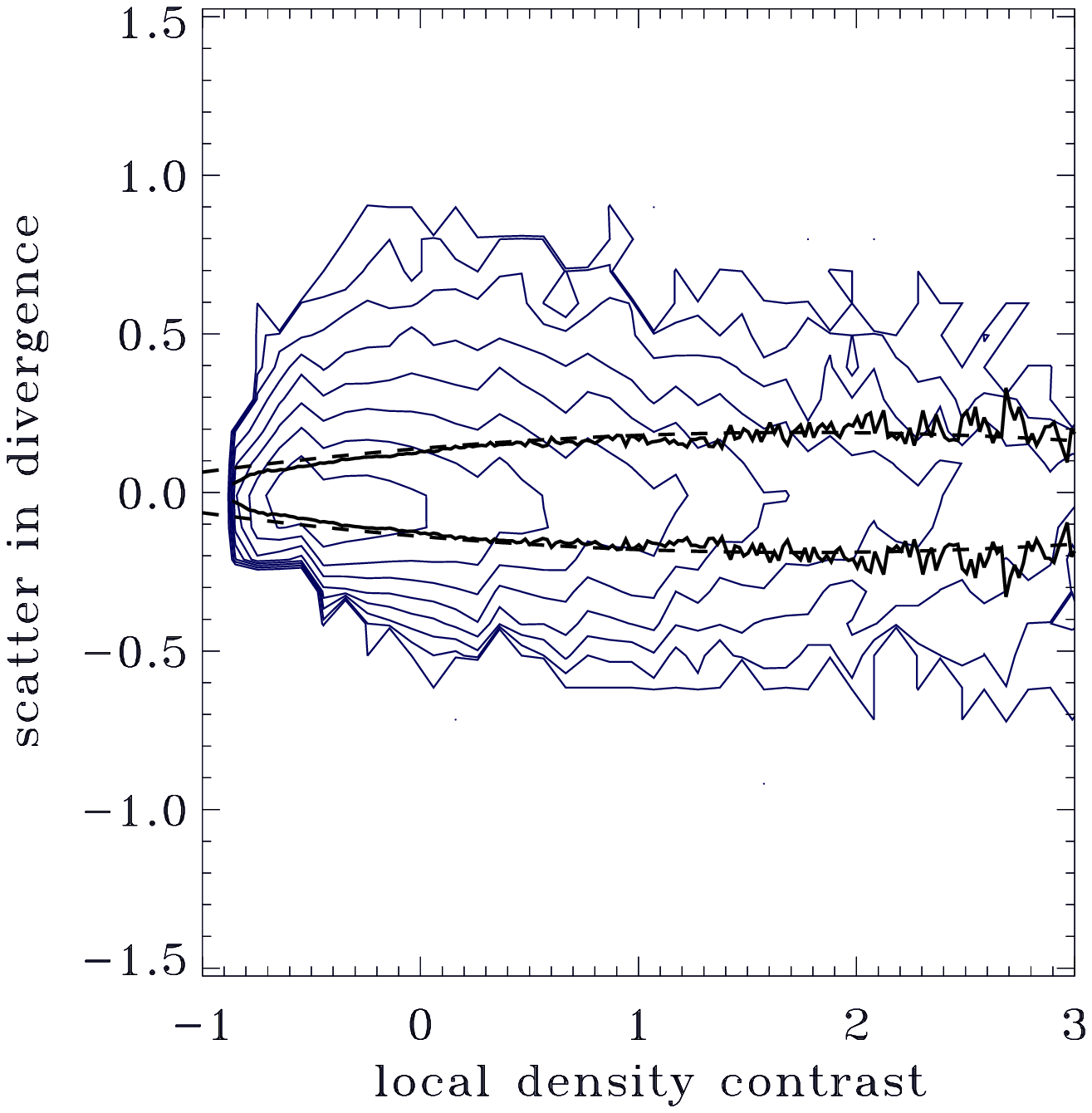,width=7.cm} 
  \end{center}
\caption{{The contour plots show the scatter of the density as a
function of the (minus) local divergence (top panel) and of the local
divergence as a function of the density (bottom panel). 
Thick (jagged) solid lines show the $\ste$ (top panel) and $\sde$
(bottom panel) obtained from numerical simulations and compared against
analytical fits (dashed lines) proposed in the text (see
Eqn. (\ref{sde}) and (\ref{ste}))
Note the nearly symmetric scatter around the mean in the
upper panel which is actually roughly compatible with a 
Gaussian distribution. The plots correspond to an
Einstein-de Sitter Universe and a smoothing 
scales of 15$h^{-1}$Mpc.}
}
\label{fig:SCoef}
\end{figure}
From third order PT we expect that the constrained variance of
$\de$ given $\te$ can be written as a first order polynomial in $\te$
times $\s_{\te}^4$, see equation~(\ref{stet}) with $b_2 = 0$ (for
details see Appendix~\ref{app:coeff}). This equation assumes however
that $\te \sim \s_{\te}$, the assumption which we have relaxed in this
Section. When $\te$ is arbitrary, but $\s_{\te}$ is smaller than
unity, we can still use all-order PT to predict the {\em
qualitative\/} behaviour of the constrained variance. Then it turns
out that at leading order in $\s_{\te}$, the formula for the `forward'
variance is given by $\s_{\te}^4$ times a power series in $\te$. An
object of more practical interest is the square root of the variance,
or the dispersion. We have found that a low-order polynomial provides
a satisfactory fit to the dispersion and finally we have chosen a
second-order polynomial. Specifically, we have found that
\be
\ste
\approx 0.45\ \left(1+\te+{2\over 9}\ \te^2\right)\ \sigma_{\te}^2
\label{ste}
\ee
is an accurate fitting formula for all cosmological  
models in our range of parameters.
As for the dispersion of the `inverse' relation, we have found that
\be
\sde
\approx 0.45\ (1+0.5 \de-0.05\ \de^2)\ \sigma_{\te}^2
\label{sde}
\ee
was a reliable fitting form.  For reasons that are not clearly
elucidated, this form is more robust
(against variation of the cosmological parameters), when written 
as a function of $\sigma_{\te}$ instead of $\sigma_{\delta}$.

In Fig. \ref{fig:SCoef} the fitting formulas (dashed lines) are tested
against the numerical results (solid lines) and are seen to be very
accurate. {Note that the expression (\ref{ste}) explicitly
predicts the scatter in the forward relation to vanish for $\te\to
-3/2$,} that is when $\te$ gets close to its lower bound, although
this was not required a priori in the fitting procedure. This is an
interesting result since it naturally makes the width of the
constrained distribution of $\delta$ small when its constrained
average gets close to $-1$. { (Though we would expect to obtain a
similar property for the inverse relation, the second-order expression
(\ref{sde}) fails to reproduce it. This expression however will not
appear in an analytic formula for the joint PDF.) Note also that the
fits recover the scaling of the dispersions with $\sigma_{\te}$
predicted by PT. Finally, the values of the coefficients of the
constant terms in expressions~(\ref{ste}) and~(\ref{sde}) are
identical, as predicted by PT. On the other hand, they are {\em not\/}
equal to the PT value. Comparing equation~(\ref{stet}) with
equation~(\ref{ste}) we see that the value of the parameter $b_0$
derived from N-body is given by $(0.45)^2 \simeq 0.2$, while, for the
spectral index $n=-1.5$ that was used, PT predicts about 0.05 (see
Table~\ref{taba1}). Once again, the PT results that involve loop terms
fail.  }

\subsection{A description of the joint PDF}
\label{sub:PDF}
\begin{figure}
  \begin{center}
    \psfig{figure=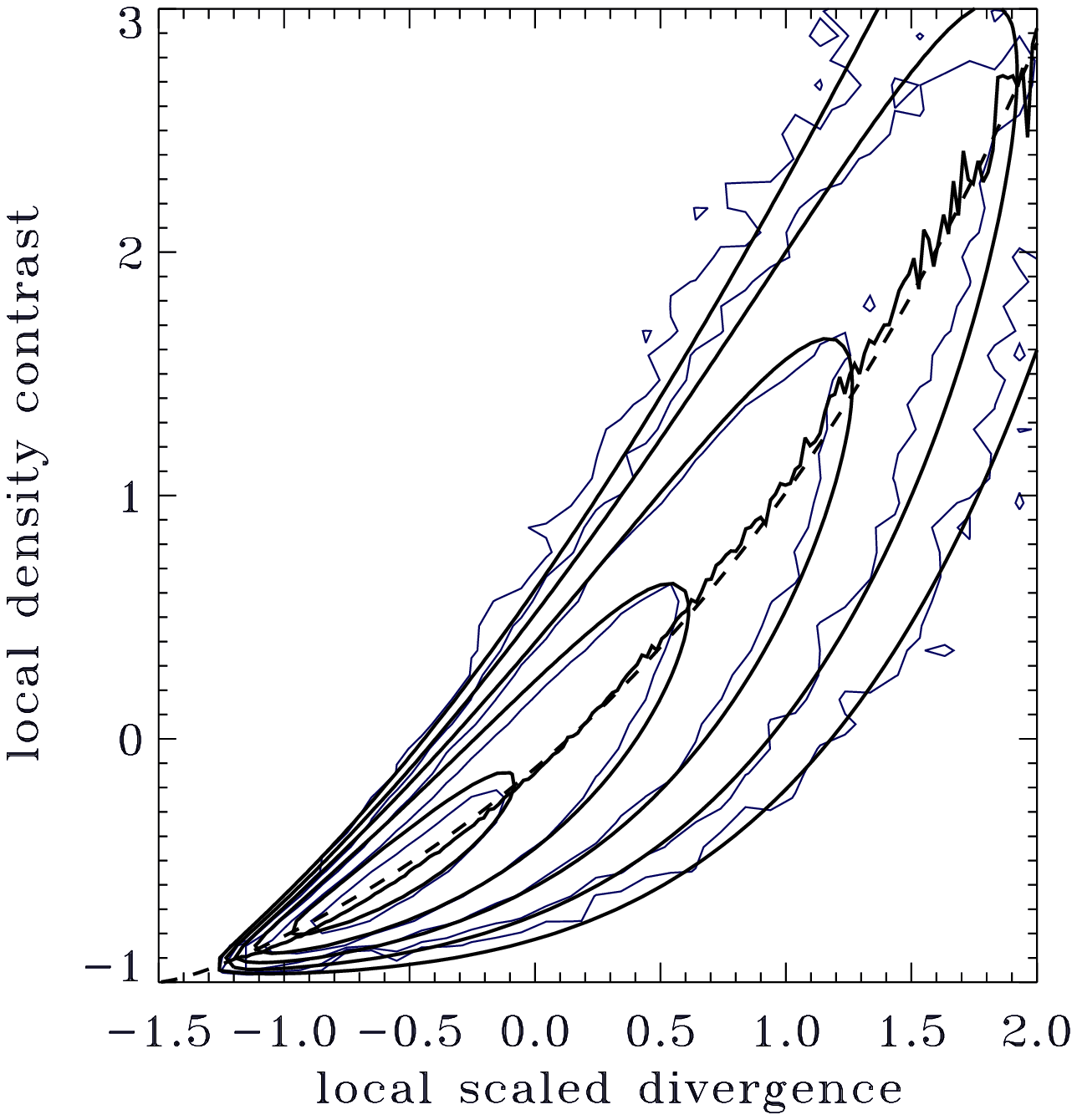,width=7.cm} 
    \psfig{figure=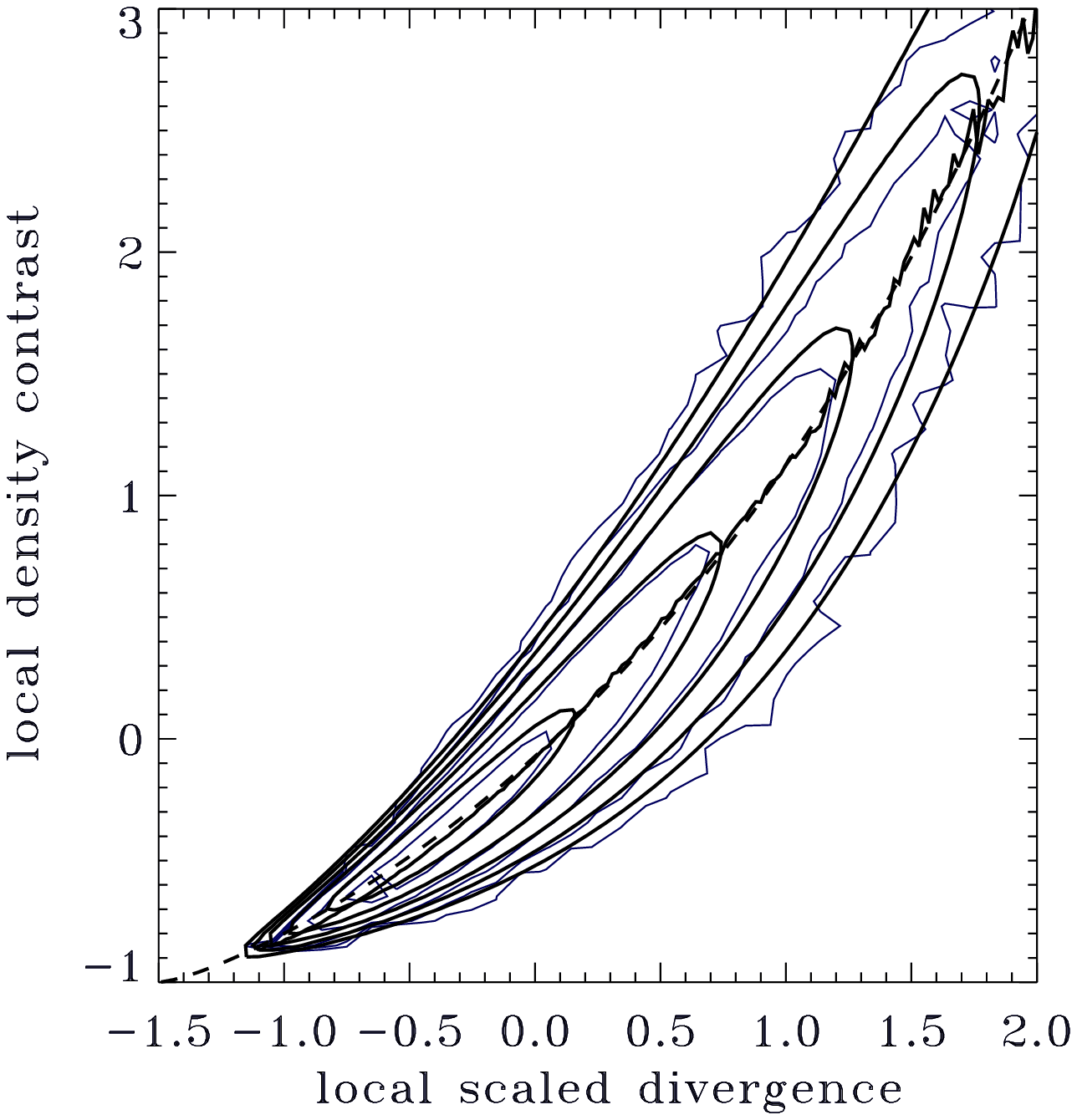,width=7.cm} 
    \psfig{figure=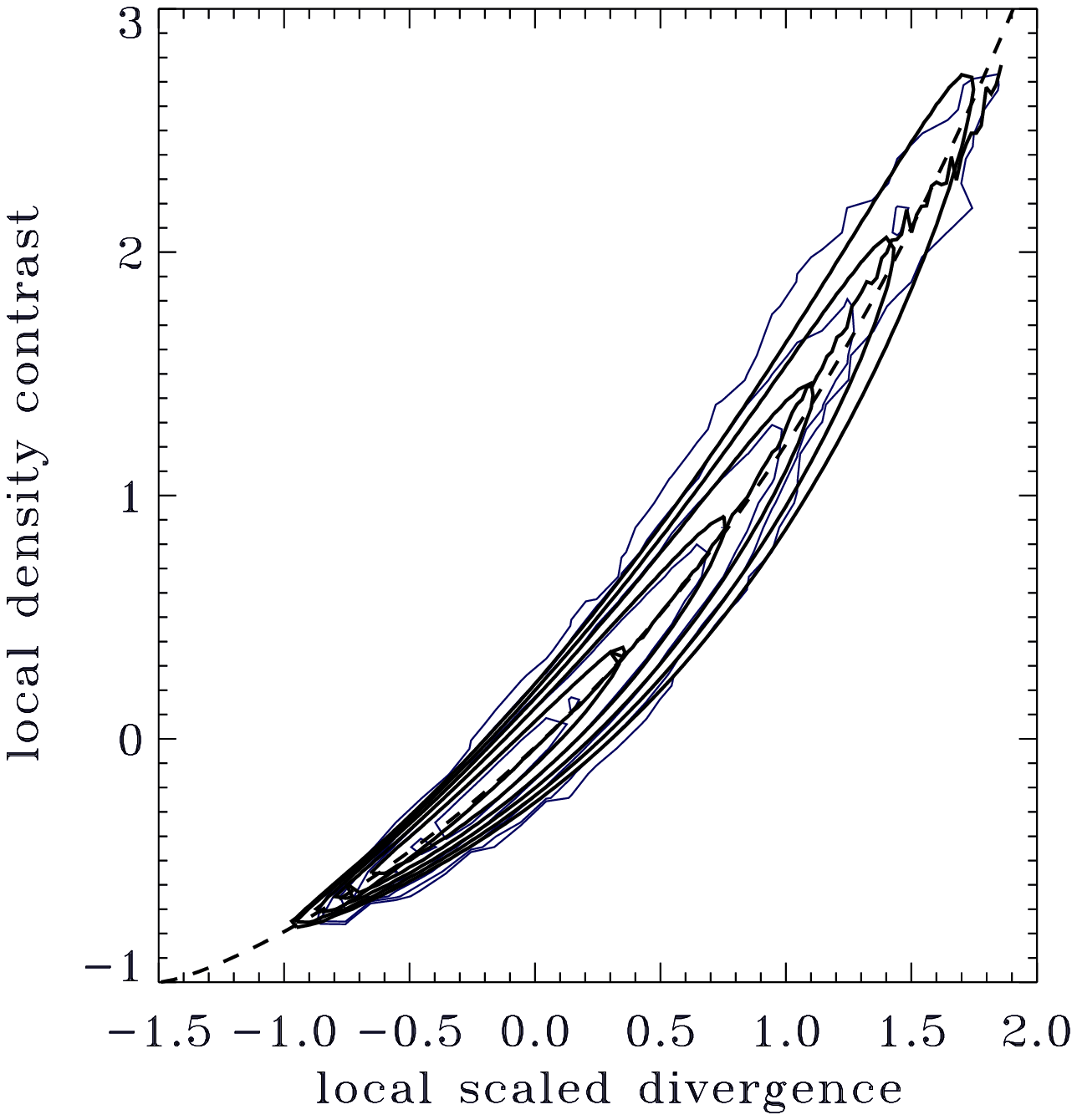,width=7.cm} 
  \end{center}
\vskip -1.truecm
\caption{Comparison of the isodensity contours
of the joint PDF obtained numerically
{(thin solid lines) and using the phenomenological formula 
(thick solid lines) proposed in the text (Eqn. (\ref{eqn:pdf2dim})).
Thicker lines are the constrained mean of $\delta$ from
numerical results (solid jagged line) and from our fit (dashed line).
The three panels correspond to three different smoothing 
scales, 10$h^{-1}$Mpc (top), 15$h^{-1}$Mpc (middle) and 25 $h^{-1}$Mpc 
(bottom) for the Einstein-de Sitter Universe.}
}
\label{fig:SPlotFlat}
\end{figure}

\begin{figure}
  \begin{center}
    \psfig{figure=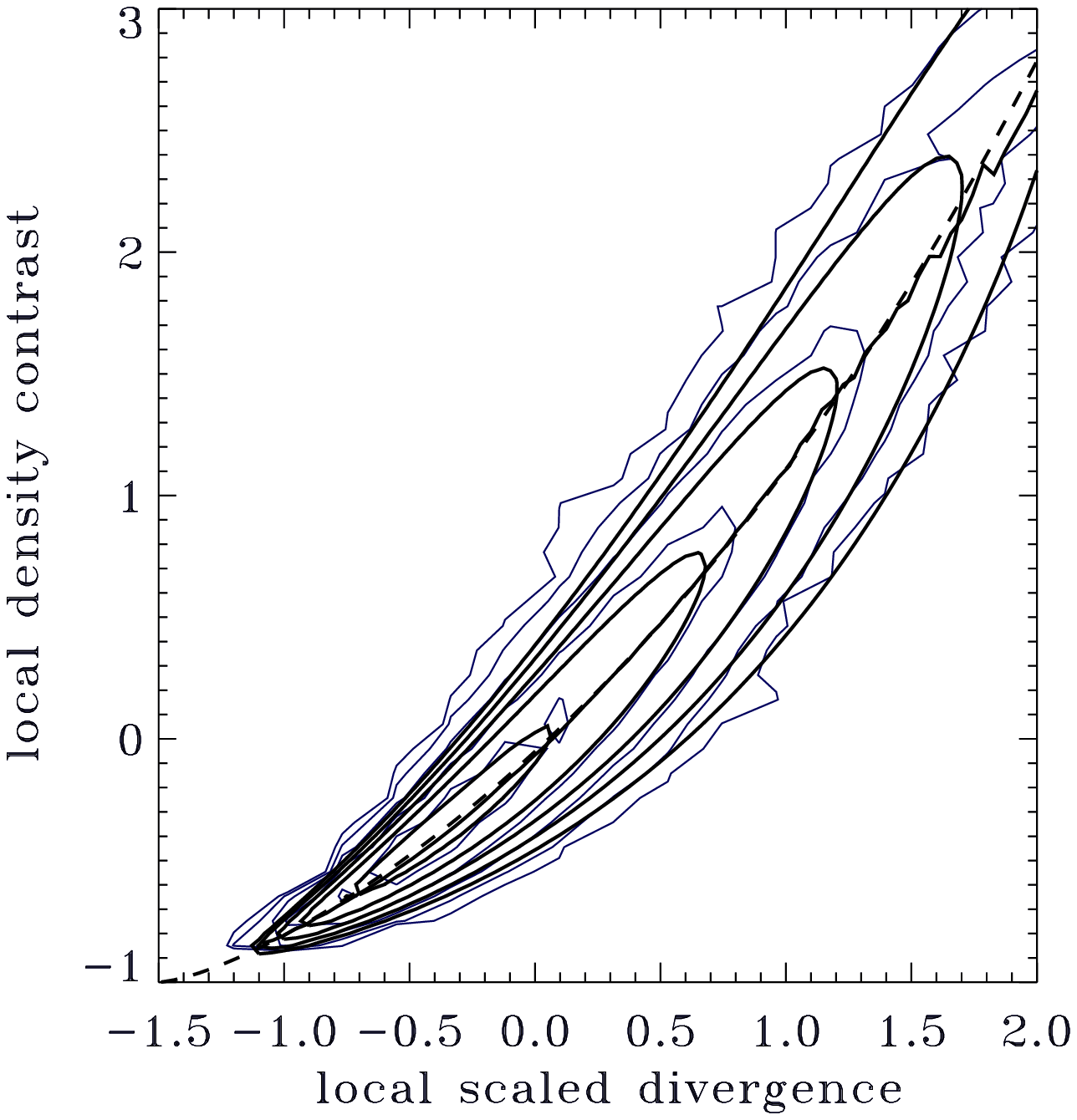,width=7.cm} 
    \psfig{figure=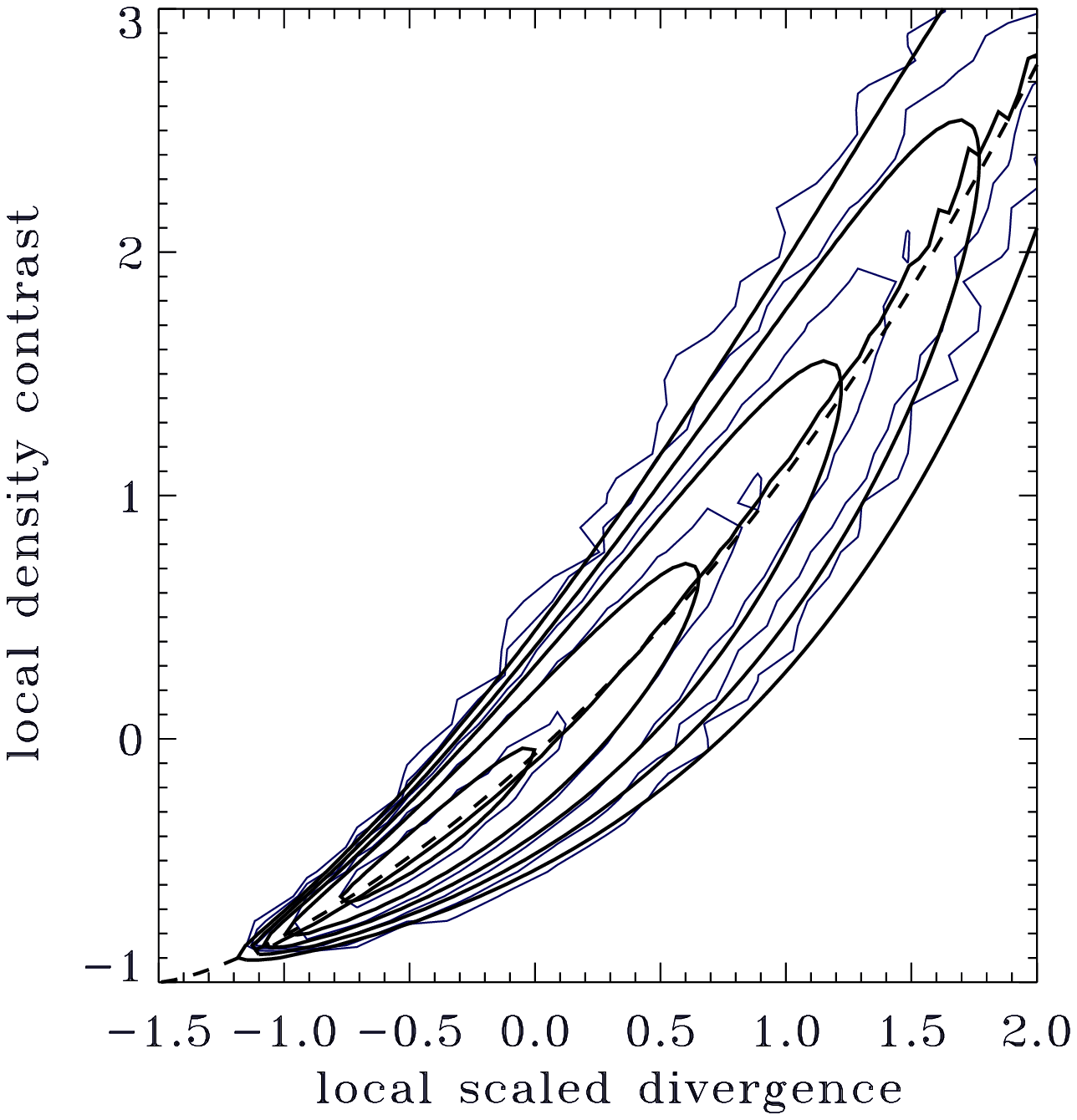,width=7.cm} 
  \end{center}
\caption{{Same as figure \ref{fig:SPlotFlat} but for other cosmologies: 
an open universe with $\Omega=0.3$ and $\lambda=0$ (top panel), and
a zero curvature Universe with $\Omega=0.3$ and $\lambda=0.7$ 
(bottom panel). A smoothing scale is equal to 15$h^{-1}$Mpc. }
}
\label{fig:SPlotOpen}
\end{figure}

{Our goal in this paper is to construct a generalization of
Equation~(\ref{Plin}) applicable in the mildly non-linear regime.}
A natural approach would be to rebuild the joint PDF from the
knowledge of the various moments. The mathematical context that
allows to do such a reconstruction is known. It is called 
the Edgeworth expansion (see Bernardeau \& Kofman 1995 for
a derivation of the Edgeworth expansion from the cumulants
and Appendix \ref{app:edge}). 
As is stressed in Appendix~\ref{app:edge}, in our case {this
expansion (or any of a similar kind) is unfortunately inapplicable.}
The reason is that at leading order the cross-correlation
matrix of $\de$ and $\te$ is singular. Therefore there exists 
a linear combination of both variables (namely $\de-\te$) which 
has a vanishing variance and the one-point PDF of which
is not necessarily close
to a Gaussian distribution even for arbitrarily small $\sigma$
(although it converges to a Dirac delta function when $\s\to 0$).

The construction of the joint PDF must then rely 
on further assumptions. We assume here that the conditional distribution
function of the density for any (fixed) value of $\te$ is Gaussian. 
This assumption is supported by the properties observed
in Fig.\ref{fig:SCoef}. It appears that a similar 
assumption for the reverse relation would have been less accurate. 
The construction of the whole joint PDF can then be done from
the knowledge of the constrained mean, the constrained scatter and
the one-point PDF of $\te$. For the later problem we took
advantage of the existence of a very accurate analytical 
fit\footnote{This PDF has been derived strictly speaking
for a $n=-1$ power spectrum.}
proposed by Bernardeau (1994b),
\ba
&&\hspace{-.7cm}
P_{\te}(\te)\der\te={[(2\kappa-1)/\kappa^{1/2}+
(\lambda-1)/\lambda^{1/2}]^{-3/2}
\over \kappa^{3/4}\ (2\pi)^{1/2}\ \s_{\te}}
\times \nonumber\\
&&\hspace{-.5cm}
\exp\left(-{\te^2\over 2\lambda\s_{\te}^2}\right)\ \der\te
\ea
with
\be
\lambda=1+{2\te\over 3},\ \ \kappa=1+{\te^2\over9\,\lambda}. 
\ee
This distribution has been found to be very accurate
in describing the PDF for the Einstein-de Sitter
Universe
 (see Bernardeau \& van de Weygaert 1996) and also
for other {Friedman-Robertson-Walker} backgrounds (Bernardeau
\etal 1997) if  the scaled divergence is used.

The {proposed expression for the} joint PDF is then given by 
\be \label{eqn:pdf2dim}
P(\de,\te)\ \der\te\ \der\de=
\exp\left[-{\left(\de-\lan\de\ran_{\te}\right)^2\over 2\ste^2}\right]
P_{\te}(\te)\,{\der\te\,\der\de\over \sqrt{2\pi\,\ste^2}}.
\ee

{As a formula for the constrained variance $\ste^2$ we adopt the
expression~(\ref{ste}). As a formula for the constrained mean we might
adopt a third-order polynomial in $\te$ with the coefficients fitted
in Subsection~\ref{sub:PTver}. However, the coefficients depend weakly
on the smoothing scale (equivalently, on $\s_{\te}$) so we should then
model their $\s_{\te}$-dependence. Furthermore, and more importantly,
as we stated earlier, in this Section we study the DVDR including its
behaviour in the tails of the joint PDF. A third-order polynomial was
overall a good fit to the data, but it did not satisfy a specific
constraint in the negative tail. Namely, for $\te \to -3/2$,
$\lan\de\ran_{\te}$ should converge to $-1$. In other words, the
maximal dimensionless expansion of voids is $-3/2$ (see Bernardeau
\etal 1997 for details). Note that the simplified formula~(\ref{B92})
naturally satisfies this constraint. This formula is approximately
valid in the limit $\s_{\te} \to 0$.
it includes contributions from all orders in PT.} We will therefore
modify it, {\em phenomenologically\/} incorporating the
$\s_{\te}$-dependence.

The formula we have found to be quite accurate for the whole range of
values of $\de$ (see Figs. \ref{fig:SPlotFlat}-\ref{fig:SPlotOpen}) is

\be \label{eqn:avfit}
\lan\de\ran_{\te}=\beta(\s_{\te})
\left(1+{\te\over \gamma}\right)^{\gamma}-1,
\ee
where $\beta$ is such that
\be
\int_{-1}^{\infty}\lan\de\ran_{\te}\ P_{\te}(\te)\ \der\te=0\,,
\ee
so it depends on the variance. This formula is relatively simple
because the $\s_{\te}$-dependence factors from the
$\te$-dependence. For the fit to be accurate also in the high-density
tail, we have slightly modified the exponent [$\gamma = 3/2$ in the
expression~(\ref{B92})] to

\be
\label{eq:gamma}
\gamma = {3\over2}+ \eps \,\left[\te+{3\over2}\ f(\Omega)\right]
\,,
\ee
where 
\be \eps \simeq 0.3 \,.
\ee
The normalization constraint leads to a value of $\beta$ given
approximately by 
\be
\beta \simeq 1 - \left[0.20+0.06 f(\Omega,\Lambda)\right] \s_{\te}^2,
\ee
for $\s_{\te}$ below 0.8. The $f(\Omega,\Lambda)$ dependence exhibited here
reflects the slight dependence of $a_2$ and $r_2$ on $\Omega$
that is found numerically. The effective coefficient $a_2$
can be perturbatively identified from the relation $\beta\simeq 1 - a_2
\s_{\te}^2$; it varies from $a_2=0.23$ for the low $\Omega$ models to
$a_2=0.26$ for the Einstein-de Sitter Universe.

This description
reveals very accurate in describing the overall shape of the
joint PDF in various regimes, for different values
of the variance and different cosmologies {as demonstrated in Figs.
\ref{fig:SPlotFlat} 
and \ref{fig:SPlotOpen} respectively.

Equation (\ref{eqn:pdf2dim}) together with (\ref{eqn:avfit}) and  (\ref{ste})
constitutes the main
result of this paper.}

\section{Towards a separate determination of the bias and $\Omega$}
\label{sub:bias}

The description of the joint distribution of $\de$ and $\te$ {discussed
in the last Section} has been derived for the matter density
distribution assuming the matter density
field is directly accessible to observations. {In case of biases
between the galaxy field and the matter density field the quantitative
results we present are not valid for the {\em galaxy}-density versus
velocity-divergence relation. However, most of the qualitative
features of the approach we propose here is expected to remain correct.
In this Section we consider some of the new quantitative features
arising if a matter-galaxy bias is present.}

Let us assume that it is possible to expand the local galaxy 
density field, $\delta^{\rm gal.}$ in terms of the {\sl
initial} matter density field,
\be
\delg(\bfx)=\delg_1(\bfx)+\delg_2(\bfx)+\ldots
\label{dgexpand}
\ee
We assume then that, from a PT point of view, the 
relation between $\delg$ and $\de$ is similar to that between
$\de$ and $\theta$. As a consequence,
the linear bias, that is the bias parameter at large scale,
is given by,
\be
\left.b^{\rm gal.}\right.^2={\lan\left.\delg_1\right.^2\ran
\over\lan\delta_1^2\ran},
\label{eq:bias}
\ee
or equivalently by,
\be
b^{\rm gal.}\equiv{\lan\delg_1\ \delta_1\ran\over\lan\delta_1^2\ran}.
\label{eq:Francis_cc}
\ee
This identification is possible
since the stochasticity is assumed to be negligible at linear order 
that is,
\be
\delg_1(\bfx)=b^{\rm gal}\delta_1(\bfx),
\ee
where $b^{\rm gal}$ is a mere number.
This assumption seems rather abrupt, but we think
that it is actually completely natural. 
{}
Indeed it does not mean
at all that the final stochasticity will vanish. It simply means
that the so called $r$ parameter ($r=\lan\delg\,\delta\ran/
\lan\left.\delg\right.^2\ran^{1/2}\,
\lan\delta^2\ran^{1/2}$) is {\it unity at very large scale}
when only the linear order plays a role. This is indeed 
observed in the numerical experiments of Blanton et al. (1999).
The second order term in Eq. (\ref{dgexpand}) however is
not assumed to be simply proportional to the square of the
local linear density contrast but may contain non-local terms
(as for the matter density contrast) or intrinsic
stochasticity. Our assumption is even weaker than what is obtained
in explicit models of biasing like BBKS for which  the
$r$ parameter is arbitrarily close to unity\footnote{It is unity in
the limit the size of the objects is much smaller than the 
scale at which the correlation coefficients are calculated.} 
(Mo \& White 1996). 

For the fields $\delg$ and $\theta$ (here we reintroduce the
observable velocity divergence field $\theta$) the forward
constrained average reads,
\ba
&&\hspace{-.7cm}
\lan\delg\ran_{\vert_{\theta}}={\lan\delg\ \theta\ran\over
\lan\theta^2\ran}\,\theta+\nonumber\\
&&\hspace{-.5cm}
\left({\lan\delg\ \theta^2\ran\over\lan\theta^2\ran^2}-
{\lan\delg\theta\ran\ \lan\theta^3\ran\over\lan\theta^2\ran^3}
\right)\ {\theta^2-\lan\theta^2\ran\over 2}+\ldots
\ea
{}
It follows from the writing of the first moment of
the joint distribution function of $\delg$ and $\theta$ (Eq. \ref{eq:meanmu})
which implies,
\ba
&&\hspace{-.7cm}
\lan\delg\ran_{\vert_{\theta}}=\nonumber\\
&&\hspace{-.5cm}
\left(-\lan\delg\theta\ran{\der\over\der\theta}P(\theta)+
{\lan\delg\theta^2\ran\over2}{\der^2\over\der\theta^2}P(\theta)+\ldots\right)
{1\over P(\theta)}
\ea
with, using the Edgeworth expansion\footnote{This expression is valid
only when the distributions of the two quantities are weakly
non-Gaussian, 
i.e., if $\lan\left.\delg\right.^p\theta^q\ran\sim\sigma^{2(p+q-1)}_{\theta}$.},
\begin{eqnarray}
P(\theta)&=&{1\over \sqrt{2\pi}\sigma_{\theta}}
\exp\left(-{\theta^2\over2\sigma_{\theta}^2}\right)\times\nonumber\\
&&\ \ \ \left[
1+{\lan\theta^3\ran\over6\,\sigma^3_{\theta}}
\left({\theta^3\over\sigma^3_{\theta}}-3{\theta\over\sigma_{\theta}}
\right)+\ldots\right].
\end{eqnarray}

Each ensemble average term appearing in this expression can be written
in terms of $f(\Omega,\Lambda)$, $b^{\rm gal.}$, $\tilde{T}_3$ 
and observable quantities. The second moments read,
\ba
\lan\delg\,\theta\ran&=&\lan\delg_1\,\theta_1\ran=b^{\rm gal.}\,
f(\Omega)\,\lan\te_1^2\ran,\\
\lan\theta^2\ran&=&f^2(\Omega)\,\lan\te_1^2\ran.
\ea
The third moment is given by, 
\ba
\lan\theta^3\ran&=&-f^3(\Omega)\,\tilde{T}_3\,\lan\te_1^2\ran^2,
\ea
where $\tilde{T}_3$ is a skewness parameter of the {\it scaled}
divergence. This is a known quantity from Perturbation
Theory that is independent of the cosmological parameters
(with a slight dependence on the power spectrum index),
\be
\tilde{T}_3={26\over 7}-(n+3).
\ee
In a perturbative expansion we have $\lan\delg\ \theta^2\ran\approx
\lan\delg_2\ \theta_1^2\ran+2\lan\delg_1\ \theta_1\theta_2\ran$
and replacing respectively $\theta_1$ by 
$-f(\Omega)/b^{\rm gal.}\,\delg_1$ 
and $\delg_1$ by $b^{\rm gal.}\te_1$
in this expression, we get 
\ba
\lan\delg\ \theta^2\ran&=&\left({1\over 3}\,S_3^{\rm gal.}\
f^2(\Omega)\,\left.b^{\rm gal.}\right.^2+\right.
\nonumber\\&&
\ \ \ \ \ \ \left.{2\over3}\,\tilde{T}_3
\,f^2(\Omega)\,b^{\rm gal.}\right)\lan\te_1^2\ran^2,
\ea
where $S_3^{\rm gal.}$ is the large--scale limit of the
galaxy skewness,
\be
S_3^{\rm gal.}={\lan\left.\delg\right.^3\ran\over
\lan\left.\delg\right.^2\ran^2}.
\ee
This quantity can be measured directly in galaxy catalogues.
The forward relation then reads,
\ba
&&\hspace{-.7cm}
\lan\delg\ran_{\vert_{\theta}}=-{b^{\rm gal.}\over f(\Omega)}\theta+
\nonumber\\
&&\hspace{-.5cm}
{\left(b^{\rm gal.} \ S_3^{\rm gal.} - \tilde{T}_3
\right) \over 6}
{b^{\rm gal.} \over f^2(\Omega)} 
\left( \theta^2-\lan\theta^2\ran \right) +\ldots
\label{biasesf}
\ea
and the inverse relation can be easily obtained either from
direct calculations or straightforwardly
from the forward relation, 
\ba
&&\hspace{-.7cm}
\lan\theta\ran_{\vert_{\delg}}=-{f(\Omega)\over b^{\rm gal.}}\theta+
\nonumber\\&&\hspace{-.5cm}
{1\over 6}\left({f(\Omega)\over b^{\rm gal.}}\ S_3^{\rm gal.}-
{f(\Omega)\over \left.b^{\rm gal.}\right.^2}\ \tilde{T}_3
\right)\, \left(\left.\delg\right.^2-\lan\left.\delg\right.^2\ran\right)
+\ldots
\label{biasesi}\ea

We stress that these relations are exact irrespective of bias models
(linearity, scale dependence and stochasticity, with the restriction
we mention for the linear bias).

As can be seen in Eqs. (\ref{biasesf}--\ref{biasesi})
the coefficient in the quadratic terms
(either $a_2$ or $r_2$) depends on $b^{\rm gal.}$ and
$f(\Omega)$ in a combination which is different from 
the usual one, ${f(\Omega)/ b^{\rm gal.}}$. This is therefore
a potential way of removing the degeneracy between 
the $f(\Omega)$ and $b^{\rm gal.}$ parameters.

The aim of this paper is not to demonstrate the viability of such
a method with real data set. This is left for future studies
(Chodorowski, in preparation). 

\section{Conclusion}

We have presented the results of Perturbation Theory on the joint relation
between the density and the velocity divergence for a top-hat filter.
These results extend those obtained previously for a Gaussian
filter (\cl, \inv). We have obtained all quantities
that can be derived in third order perturbative calculations. 
That includes 
\begin{itemize}
\item next to leading order terms (containing so called loop terms)
for the linear coefficients $a_1$ and $r_1$ in eqs. (\ref{eq:for}, 
\ref{eq:inv}); 
\item the coefficients $a_2$, $a_3$, $r_2$ and $r_3$
describing the constrained averages expanded as third order
polynomials;
\item the coefficients $b_0$ and $s_0$
describing the constrained scatter up to the order four in $\sigma$.
These coefficients also contain loop terms.
\end{itemize}

We have then used a series of numerical simulations to check the qualitative
and quantitative behaviours predicted by these results.
We have found a good agreement with all predicted qualitative behaviours.
All PT results that are strictly at leading order (no loop terms
included) have been found to be quantitatively well reproduced. However
all the predictions involving one-loop calculations (that 
includes $a_1$, $r_1$, $b_0$ and $s_0$ coefficients) have been
found to overestimate the results. This is particularly obvious
for the linear coefficients (in Fig. \ref{fig:linCoef}) where the departure
from the horizontal lines (linear predictions) is much weaker
than the one loop order PT predictions (thin dot-dashed lines).

The second part of our work consisted in building a complete
description of a joint distribution 
of the density and local velocity divergence. 
It appears, as we pointed out, that 
it is not possible to build the joint PDF from a proper
Edgeworth expansion, as it should be natural in the regime
we are. We have thus been forced to make further assumption on the
statistical behaviour of some reduced quantities, namely that 
the constrained density is Gaussian distributed.

To achieve the joint PDF 
construction we have taken advantage of the PT results that revealed
accurate but have also 
extended them, using the scaling relations they suggested, 
with numerical fits when necessary. This is important in
particular when one wants to deal 
with rare event tails (which are not a priori properly
described by PT results).

Our final description includes nonlinearities (such as the skewness)
in both the one-point PDF of each quantity 
and in the expressions of the constrained
means. The nonlinearities, because of the nonlocal effects they
contain, also induce a stochasticity in this relation. We have described
here its generic behaviour in terms of the constrained scatter and
{eventually proposed an approximate expression for the joint PDF
providing a complete statistical description of the relation between
the local divergence and the local density in the discussed regime.}
{Though for technical
reasons we have focused our presentation on top-hat filtered fields, we
expect no significant changes for other window functions.}

{Presented results may facilitate more advanced comparisons
of the local density--velocity data of the present day survey or the
survey to be available in the coming years and provide us with
a clue towards a separate determination of biases and $\Omega$.}

{Using as a specific example the relation between matter and galaxy fields,
we have also discussed how local scalar fields employed in cosmology
might be related at scales of a cosmological interest. 
In particular, we have shown that nonlinearities
are bound to induce not only a nonlinear bias but a significant
amount of stochasticity in their respective relation even if that is not
present at the linear order.}

\section*{Acknowledgments}
We are grateful to Hugh Couchman for a copy of his AP3M code.
F.B. and M.C. thank IAP for its hospitality. This research has been
supported in part by the Polish State Committee for Scientific
Research grants No.~2.P03D.008.13 and 2.P03D.004.13, and the Jumelage
program `Astronomie France--Pologne' of CNRS/PAN.
R.S. is supported by NASA AISRP grant NAG-3941.

\appendix
\section{the joint density-divergence PDF}
\label{app:pdf}
\subsection{General expression of the joint PDF}
The joint PDF of variables $\de$ and $\te$, $P(\de,\te)$, is
given by the inverse Fourier transform of its characteristic function,
$\Phi$. The characteristic function is related to the cumulant
generating function, $\calK$, by the equation 
\be
\Phi(it,it') = \exp{[\calK(it,it')]} \,. \label{pdf2}
\ee
The cumulants, $\kappa_{pq}$, from which $\calK$ is constructed,
\be
\calK = \sum_{(p,q) \ne (0,0)}^\infty  \f{\kappa_{pq}}{p! q!}
(it)^p (it')^q \,,  \label{pdf3}
\ee
are given by the {\em connected} part of the joint moments
\be
\kappa_{pq} = \lan \de^p \te^q \ran_{c} \,.
\label{pdf4}
\ee
The cumulants are a convenient measure of non-Gaussianity, since for a
(bivariate) Gaussian distribution they all vanish for $p + q \ge 3$.
 
The variables $\de$ and $\te$ have zero mean, hence $\kappa_{10} =
\kappa_{01} = 0$. Then from equations~(\ref{pdf2}) and~(\ref{pdf3}) we have
\begin{eqnarray}
P(\de,\te) = \f{1}{(2\pi)^2} \int\!\!\int \exp\big[ -it\ \de
-it'\ \te - \nonumber \\
{\textstyle \f{1}{2}} (\kappa_{20} t^2 + 
2\kappa_{11} tt' + \kappa_{02} t'^2) \big]
\nonumber \\
\ \times \exp{\left[\sum_{p+q \ge 3}^\infty  \f{\kappa_{pq}}{p! q!}
(it)^p (it')^q \right]} \, {\rm d}t {\rm d}t' \,.  
\label{pdf5}
\end{eqnarray}
Writing $s = \kappa_{20}^{1/2} t$, $s' =
\kappa_{02}^{1/2} t'$, introducing the standard variables
\be
\mu = \de/\kappa_{20}^{1/2} \qquad \hbox{and} \qquad
\nu = \te/\kappa_{02}^{1/2} \,,
\label{pdf6}
\ee
and the standard cumulants
\be
\lam_{pq} = \f{\kappa_{pq}}{\kappa_{20}^{p/2} \kappa_{02}^{q/2} }
\,, 
\label{pdf7}
\ee
yields 
\begin{eqnarray}
&&\hspace{-.7cm}
P(\mu,\nu) = \nonumber \\
&&\hspace{-.7cm}
\f{1}{(2\pi)^2} 
\int\!\!\int \exp{[ -i(\mu s + \nu s') - {\textstyle \f{1}{2}} 
(s^2 + 2\lam_{11} ss' + s'^2) ]}
\times\nonumber \\
&&\hspace{-.7cm}
 \exp{\left[\sum_{p+q \ge 3}^\infty  \f{\lam_{pq}}{p! q!}
(is)^p (is')^q \right]} \, {\rm d}s {\rm d}s' \,.  
\label{pdf8}
\end{eqnarray}
\subsection{Conditional probabilities and conditional moments}
The conditional probability, $P(\mu\vert\nu)$, reads,
\be
P(\mu\vert\nu)\equiv {P(\mu,\nu)/ P(\nu)}.
\ee
The conditional moments are then the moments of $P(\mu\vert\nu)$. So
the constrained average of $\mu$ is
\be
\lan\mu\ran_{\vert \nu}={\int \der\mu\ \mu\ P(\mu,\nu)/ P(\nu)}
\ee
It is possible to express this result in term of the cross-correlation
coefficients. Using the general expression of the distribution
in terms of the cumulant generating function we indeed have,
\ba
&&\hspace{-.7cm}
\lan\mu\ran_{\vert \nu}={1\over (2\,\pi)^2\ P(\nu)}\times\nonumber\\
&&\hspace{-.7cm}
{\int \der\mu\ \der s\ \der s'\ \mu\exp[-i(\mu\
s+\nu\ s')+\Lambda(s,s')]},
\ea
with
\be
\Lambda(s,s')= 
\sum_{p+q \ge 2}^\infty  \f{\lam_{pq}}{p! q!}\,(is)^p\,(is')^q \,.
\ee
After some mathematics, we have,
\be
\lan\mu\ran_{\vert \nu}={1\over 2\pi\,P(\nu)}{\int \der s'\ 
{\partial \Lambda\over\partial s}(0,s')\ \exp[-i\nu\,s'+ \Lambda(0,s')]}
\ee
which, when re-expressed in terms of $P(\nu)$, gives 
\be
\lan\mu\ran_{\vert \nu}={1\over P(\nu)}\,\sum_{q \ge 1}
(-1)^q \f{\lam_{1,q}}{q!} 
{\left({\der\over\der\nu}\right)^q\,P(\nu)}.
\label{eq:meanmu}
\ee
The higher order moments can be calculated in a similar way.

\section{constrained averages in Perturbation theory}

In perturbation theory the local density $\de$ and the local
divergence $\te$ can be expanded in terms of the linear solution.
We write these expansions,
\be
\de=\de_1+\de_2+\de_3+\ldots
\ee
and
\be
\te=\te_1+\te_2+\te_3+\ldots
\ee
Statistical quantities of interest will then be build out of connected
moments involving any order term of these expansion such as
$\lan\de_i\dots\de_j\,\te_k\dots\te_l\ran_c$. Taking advantage of
this expansion it is easy to see that, in the weakly nonlinear regime,
the standard cumulants, { defined in equation~(\ref{pdf7}),} obey
for $p + q \ge 2$ the following scaling hierarchy (Fry 1984,
Bernardeau 1992a),
\be
\lam_{pq} = S_{pq}\ \s^{p+q-2} + {\cal O}(\s^{p+q}) 
\label{scales}
\ee 
where $\s$ is the linear variance of $\de$ or, equivalently, of $\te$
(we recall that $\de_1 = \te_1$). {(The coefficients $S_{pq}$ in
the above equation are called `the hierarchical parameters'. They are
joint skewness, joint kurtosis, and so on.)}  As a result the series
in $\lambda_{1,q}$ in (\ref{eq:meanmu}) is equivalent to a Taylor
expansion in $\sigma$. The complete formulae can then be obtained
using the Edgeworth expansion for the distribution $P(\nu)$. We
emphasize here that the Edgeworth expansion is not just `a convenient
form' of the PDF, but a direct consequence of the hierarchy $\lam_{p}
= S_{p}\ \s^{p-2}$, obeyed by the standard cumulants in the weakly
nonlinear regime. The derivation of the third-order Edgeworth
expansion for a single variable is well-known (e.g.,
Longuet-Higgins~1963, Bernardeau \& Kofman 1995, Juszkiewicz \etal
1995), \ba &&\hspace{-.7cm} P(\nu) = \frac{1}{(2 \pi)^{1/2}}
\exp{(-\nu^{2}/2)} \times\nonumber\\ &&\hspace{-.7cm} \left[ 1 +
{T_{3}\over 6} \sigma H_{3}(\nu) + {T_{4}\over 24} \sigma^{2}
H_{4}(\nu) + {T_{3}^{2}\over 72} \sigma^{2} H_{6}(\nu) + \ldots
\right] \,,
\label{egde1d}
\ea
where $T_p\equiv S_{0q}$ and $H_n$ are Hermite polynomials. 
This formula is written here up to third order in $\s$ but
could be generalized to any order.
The conditional moments of $\mu$ can then be calculated
straightforwardly. We express them in terms of
quantities that are accessible to PT calculations.
The first moment reads, 
\ba
&&\hspace{-.7cm}
\lan \mu \ran_{\vert_\nu} = \left(1 - \f{\alpha \s^2}{2}\right) \nu + 
\f{(S_3 - T_3) \s}{6} (\nu^2 - 1) +\nonumber\\
&&\hspace{-.7cm}
\f{\Sigma_4 \s^2}{6} H_3(\nu) 
- \f{(S_3 - T_3)\ T_{3} \s^2}{6} (\nu^3 - 2\nu) \,,
\label{mmu}
\ea
where $S_3$ is the skewness (properly normalized
third order moment) of the density field,
\be
S_{3} = \f{3 \lan \de_1^2 \de_2 \ran_c}{\s^4}\,,
\label{t3}
\ee
and $T_3$ is the skewness of the scaled velocity divergence defined in
an analogous way.  Dividing by $\sigma^4$ results in the independence
of $S_{3}$ and $T_{3}$ on the normalization of the spectrum and, in
the case of power-law spectra, smoothing scale. The remaining
quantities are 
\be
\alpha \equiv \frac{\langle (\delta_2 - \theta_2)^2 \rangle}{\sigma^4} 
\label{alpha1}
\ee
and 
\begin{equation}    \label{Sig4}
    \Sigma_{4} = \frac{3 \langle \delta_{1}^{2} \delta_{2} \theta_{2} 
    \rangle_c - 3 \langle \theta_{1}^{2} \theta_{2}^{2} \rangle_c + 
    \langle \delta_{1}^{3} \delta_{3} \rangle_c - \langle \theta_{1}^{3} 
    \theta_{3} \rangle_c}{\sigma^{6}} . 
\end{equation} 
Expression (\ref{mmu}) is of course in agreement with the result of \cl. 
By equation~(\ref{pdf12}), the first term in this expression is
equal to $r_{} \nu$, resembling the case of purely Gaussian variables,
where $\lan \mu \ran_{\vert_\nu} = r_{} \nu$. However, nonlinearities of the
fields in question induce corrections to the mean trend that contain
also other terms linear in $\nu$.
The constrained variance, 
$\s^2_{{\mu}\vert{\nu}}\equiv \lan\mu^2\ran_{\vert_\nu} - 
\lan\mu\ran_{\vert_\nu}^2$, is given by,
\ba
&&\hspace{-.7cm}
\s^2_{{\mu}\vert{\nu}} = 
\left( \alpha - \f{\Delta K}{2} + \f{(S_3 - T_3)^2}{18} \right) \s^2
+\nonumber\\
&&\hspace{-.7cm}
\left( \f{\Delta K}{2} - \f{(S_3 - T_3)^2}{9} \right) \s^2 \nu^2 \,,
\label{smu}
\ea
with,
\be
\Delta K = \f{\lan \te_1^2 (\de_2 - \te_2)^2 \ran_c}{\s^6},
\label{DK}
\ee
and in agreement with the leading-order formula in \inv. 

The above formulas are given for standard variables, while for
practical applications we would like to have them for the `physical'
ones, $\de$ and $\te$. The needed transformation involves the
variances of $\de$ and $\te$ which are equal only at linear order. 
Taking this into account (see \cl\ for details) and rearranging the terms
in equation~(\ref{mmu}) yields expression~(\ref{eq:for}) for the mean
density given the velocity divergence with the coefficients
\begin{eqnarray} 
    a_1 &=& 1 + \left[\Sigma_2 + \frac{(S_{3} - T_{3})\, 
    T_{3} }{3} - \frac{\Sigma_4}{2} \right] \sigma^2 \,,
    \label{a1} \\ 
    a_2 &=& \frac{S_{3} - T_{3}}{6} \,,
    \label{a2} \\ 
    a_3 &=& \frac{\Sigma_4 - (S_{3} - T_{3}) T_{3} }{6}
    \,.
    \label{a3} 
\end{eqnarray}
The quantities $\Sigma_2$ and $\Sigma_4$ are given by
\begin{equation}     \label{Sig2}
    \Sigma_{2} = \frac{ \langle \delta_{2} \theta_{2} \rangle_c - \langle
    \theta_{2}^{2} \rangle_c + \langle \delta_{1} \delta_{3} \rangle_c -
    \langle \theta_{1} \theta_{3} \rangle_c}{\sigma^{4}}
\end{equation}
and by Eq. (\ref{Sig4}).
The quantity $\Sigma_2$ is built of terms similar to the properly
normalized weakly nonlinear corrections to the variance of the density
and velocity divergence fields (Scoccimarro \& Frieman 1996, \L okas
\etal 1996).
In the case of density, for example, we have to the leading
order
\begin{equation} \label{sde2} 
\frac{\langle \delta^2 \rangle - \sigma^2}{\sigma^4} = \frac{\langle 
    \delta_{2}^{2} \rangle + 2 \langle \delta_{1} \delta_{3} 
    \rangle}{\sigma^{4}} \,.
\end{equation}
On the other hand, $\Sigma_4$ is composed of quantities similar to the
properly normalized kurtosis which for the density field, to the
leading order is (Bernardeau 1994a; {\L}okas et al. 1995)
\begin{equation}        \label{s4}
    S_{4} = \frac{6 \langle \delta_{1}^{2} \delta_{2}^{2} \rangle
    + 4 \langle \delta_{1}^{3} \delta_{3} \rangle}{\sigma^{6}} \,.
\end{equation}
The inverse relation~(\ref{eq:inv}), giving mean $\theta$ when
$\delta$ is known, is obtained immediately from equations
(\ref{eq:for}), (\ref{Sig4}) and (\ref{a1})--(\ref{Sig2}) via
exchange of variables $\de$ and $\te$. We have thus
\begin{eqnarray} 
    r_1 &=& 1 + \left[ \Sigma_2' + \frac{(T_{3} - S_{3})\, 
    S_{3} }{3} - \frac{\Sigma_4'}{2} \right] \sigma^2 \,,
    \label{r1} \\ 
    r_2 &=& \frac{T_{3} - S_{3}}{6} \,,
    \label{r2} \\ 
    r_3 &=& \frac{\Sigma_4' - (T_{3} - S_{3}) S_{3}
    }{6} 
    \,,
    \label{r3} 
\end{eqnarray} 
where 
\begin{equation}    \label{Sig2p}
    \Sigma_{2}' = \frac{ \langle \theta_{2} \delta_{2} \rangle_c - \langle 
    \delta_{2}^{2} \rangle_c + \langle \theta_{1} \theta_{3} \rangle_c - 
    \langle \delta_{1} \delta_{3} \rangle_c}{\sigma^{4}}  
\end{equation} 
and 
\begin{equation}    \label{Sig4p}
    \Sigma_{4}' = \frac{3 \langle \theta_{1}^{2} \theta_{2} \delta_{2} 
    \rangle_c - 3 \langle \delta_{1}^{2} \delta_{2}^{2} \rangle_c + \langle 
    \theta_{1}^{3} \theta_{3} \rangle_c - \langle \delta_{1}^{3} \delta_{3} 
    \rangle_c}{\sigma^{6}} \,.
\end{equation}

As already stated, relations~(\ref{eq:for}) and~(\ref{eq:inv}) are not
deterministic. The scatter around the mean given by these relations 
is up to terms of the order of $\sigma^5$ given by
expressions~(\ref{stet}) and~(\ref{sdet}) respectively. [These
expressions are next-to-leading-order extensions of
formula~(\ref{smu}).] The coefficients $b_0$ and $b_2$ in
expression~(\ref{stet}) for $\ste$ are
\begin{equation}    \label{b0}
    b_0 = \frac{\langle (\delta_2 - \theta_2)^2 \rangle}{\sigma^4} - 
    \frac{\langle \theta_1^2 (\delta_2 - \theta_2)^2) \rangle}{2 \sigma^6}
    + \frac{(S_{3} - T_{3})^2}{18}
\end{equation}
and
\begin{equation}    \label{b2}
    b_2 =  \frac{\langle \theta_1^2 (\delta_2 - \theta_2)^2) \rangle}{2 
    \sigma^6} - \frac{(S_{3} - T_{3})^2}{9} \,.
\end{equation}

The coefficients $b_0$ and $b_2$ are invariants with respect to
exchange of $\de$ with $\te$. This yields equation~(\ref{b0res}). The
linear variance of the density field is given by
\begin{equation}     \label{sig2}
    \sigma^{2}=\langle \delta_{1}^{2} \rangle = D^{2}(t) \int
    \frac{{\rm d}^3 k}{(2 \pi)^{3}} \, P(k) \, W^{2}(k R)
\end{equation}
where $D(t)$ is the linear growth factor [$\delta_1({\bmath x}, t) =
D(t) \delta_1({\bmath x})$] and $P(k)$ is the linear power spectrum of
density fluctuations.

\section{Calculation of the coefficients for a top-hat filter}
\label{app:coeff}

The purpose of this section is to calculate the values of the coefficients 
$a_m$, $r_m$ and $b_m$ for the top-hat filter with Fourier representation 
\begin{equation} \label{wth}
    W(kR)= 3 \sqrt{\frac{\pi}{2}} (k R)^{-3/2} J_{3/2}(k R)
    \,,
\end{equation}
where $J_{3/2}$ is a Bessel function
\begin{equation}    \label{j32}
    J_{3/2}(x) = \sqrt{\frac{2}{\pi x}} \left( \frac{\sin x}{x} - \cos x 
    \right).
\end{equation}
Up to third order it is sufficient to consider the linear variance,
identical for both the density and the scaled velocity divergence
fields, $\s^2$.

In the following we will restrict ourselves to
the case of scale-free spectra
\begin{equation}    \label{pk}
    P(k) = C k^n, \ \ -3 \leq n \leq 1.
\end{equation}
In the case of top-hat window function and scale-free spectra the 
variance is
\begin{equation}    \label{sig2th}
    \sigma^{2} = C D^{2}(t) \,\frac{9 \Gamma[(n+3)/2] \Gamma[(1-n)/2]}
    {8 \pi^{3/2} R^{n+3} \Gamma(1-n/2) \Gamma[(5-n)/2]}. 
\end{equation}

In the calculations of the coefficients we extensively use the results
of 
Bernardeau (1994a) that led to the 
expressions for skewness and kurtosis 
for density and velocity fields for top-hat smoothing. The skewness of 
the density and the velocity divergence are then respectively
\begin{equation}    \label{s3th}
    S_{3} = \frac{34}{7} - (n+3)
\end{equation}
and
\begin{equation}    \label{t3th}
    T_{3} = \frac{26}{7} - (n+3) \,.
\end{equation}
Using equations (\ref{s3th})-(\ref{t3th}) we find
that contrary to the case of Gaussian smoothing the coefficients $a_2$
and $r_2$ do not depend on the spectral index,
\begin{equation}  \label{a2th}
    a_2 = - r_2 = \frac{4}{21} \simeq 0.190 \,.
\end{equation}
Finding $a_3$ and $r_3$ is a more demanding task. In order to
calculate $\Sigma_4$ and $\Sigma_4'$ we need to combine the results
for different kurtosis-type terms obtained by Bernardeau (1994a) and in
addition find the mixed term $\langle \delta_{1}^{2} \delta_2 \theta_2
\rangle / \sigma^6$.  Using equations~(A28) and~(A30) of Appendix A of
Bernardeau (1994a) we have
\begin{eqnarray}
    \frac{\langle \delta_1^2 \delta_2^2 \rangle}{\sigma^6} & = &
    \frac{2312}{441} - \frac{157}{63}(n+3) + \frac{5}{18} (n+3)^2 + 
    \frac{2}{9} \overline{k^2} \,, \nonumber \\
    \frac{\langle \theta_1^2 \theta_2^2 \rangle}{\sigma^6} & = &
    \frac{1352}{441} - \frac{125}{63}(n+3) + \frac{5}{18} (n+3)^2 + 
    \frac{2}{9} \overline{k^2} \,, \nonumber \\    
    \frac{\langle \delta_1^2 \delta_2 \theta_2 \rangle}{\sigma^6} & = &
    \frac{1768}{441} - \frac{47}{21}(n+3) + \frac{5}{18} (n+3)^2 + 
    \frac{2}{9} \overline{k^2} \,, \nonumber \\
    \frac{\langle \delta_1^3 \delta_3 \rangle}{\sigma^6} & = &
    \frac{682}{189} - \frac{10}{7}(n+3) + \frac{1}{6} (n+3)^2 -
    \frac{1}{3} \overline{k^2} \,, \nonumber \\
    \frac{\langle \theta_1^3 \theta_3 \rangle}{\sigma^6} & = &
    \frac{142}{63} - \frac{22}{21}(n+3) + \frac{1}{6} (n+3)^2 -
    \frac{1}{3} \overline{k^2} \,, \label{rth}
\end{eqnarray}
where
\begin{equation}  \label{ok2}
    \overline{k^2} = \frac{D^{2}(t)}{\sigma^2} \int \frac{{\rm d}^3 k}{(2 
    \pi)^{3}} \, k^2 P(k) \, W^{2}(k R) \,.
\end{equation}
The terms containing $\overline{k^2}$ cancel out when we calculate 
$\Sigma_4$,
\begin{equation}  \label{Sig4th}
    \Sigma_4 = \frac{5536}{1323} - \frac{8}{7} (n+3) \,,
\end{equation}
and by combining with our equations (\ref{s3th})-(\ref{t3th}) we finally
get the value of the coefficient $a_3$ which is similarly to $a_2$
independent of the spectral index $n$,
\begin{equation}  \label{a3th}
    a_3 = - \frac{40}{3969} \simeq -0.0101 \,.
\end{equation}
An analogous calculation yields
\begin{equation}  \label{r3th}
    r_3 = \frac{328}{3969} \simeq 0.0826 \,.
\end{equation}
The values (\ref{a2th}) and~(\ref{a3th})--(\ref{r3th}) are the same as in
the case of no smoothing, calculated by \cl\ and \inv.
The values of the coefficients $a_{1}$ and $r_{1}$ do depend on the
spectral index, their behaviour is therefore similar to the case of
Gaussian smoothing. The calculations for a top-hat filter can be
performed in a way strictly analogous to those presented for a
Gaussian smoothing in the Appendix C of \cl. The results for $a_1$ and
$r_1$ as functions of the spectral index are given in Table~\ref{taba1}.
\begin{table}
\caption{The coefficients $a_1$, $r_1$ and $b_0$ in the most
interesting range of spectral indices for power law spectra and
top-hat smoothing}
\label{taba1}
\begin{tabular}{clll} 
  $n$ & \ \ $(a_{1}-1)/\s^2$ & \ \ $(r_{1}-1)/\s^2$ & \ \ \ \ $b_0$ \\
  \hline
-2.0 & $-0.172$  & $+0.0244$ & 0.0231\\
-1.9 & $-0.134$  & $-0.0302$ & 0.0260\\
-1.8 & $-0.0850$ & $-0.0991$ & 0.0299\\
-1.7 & $-0.0217$ & $-0.190$ & 0.0350\\
-1.6 & $+0.0643$ & $-0.317$ & 0.0421\\
-1.5 & $+0.187$  & $-0.507$ & 0.0524\\
-1.4 & $+0.376$  & $-0.825$ & 0.0681\\
-1.3 & $+0.698$  & $-1.45$ & 0.0946\\
-1.2 & $+1.36$   & $-3.06$ & 0.148 \\
-1.1 & $+3.37$   & $-10.9$ & 0.309 \\
\end{tabular}
\end{table}
{}From equations~(\ref{b2}), (\ref{s3th}), (\ref{t3th}) and~(\ref{rth})
it follows straightfordwardly that
\be
b_2 = 0 \,.
\label{b2th}
\ee
Expression~(\ref{b0}) can be cast to the following form
\be
b_0 = \alpha - b_2 - 2 a_2^2 \,.
\label{b0th}
\ee
In turn, $\alpha$ (Eq. \ref{alpha1}) can be expressed as 
\be
\alpha = - \Sigma_2 - \Sigma_2' \,,
\label{alpha2}
\ee 
where $\Sigma_2$ and $\Sigma_2'$ enter definitions~(\ref{a1})
and~(\ref{r1}) of $a_1$ and $r_1$, respectively. Therefore, all the
terms determining the value of the coefficient $b_0$ have been already
computed. The resulting value of $b_0$ as a function of the spectral
index is presented in the last column of Table~\ref{taba1}. 

\section{The Edgeworth expansion for joint distribution}
\label{app:edge}

In the weakly nonlinear regime, the double series
in equation~(\ref{pdf8}) is a power series in a small parameter
$\s$. A natural idea is therefore to truncate the series at some order
$p + q = n$, similarly to when deriving the Edgeworth expansion of a {\em
one-point} PDF. One then expands the exponent in integral~(\ref{pdf8})
and retains the terms up to the order of $\s^{n - 2}$. These terms
are subsequently integrated using the following lemma:
\begin{eqnarray} 
&&\hspace{-.7cm}
\f{1}{2 \pi} 
\int\!\!\int_{- \infty}^{\infty} 
e^{[ -i(\mu s + \nu s') - {\textstyle \f{1}{2}} 
(s^2 + 2r_{} ss' + s'^2) ]} (is)^{p'} (is')^{q'} {\rm d}s {\rm d}s'=
\nonumber \\
&&\hspace{-.7cm}
(1 - r_{}^2)^{-1/2} H_{p'q'}(\mu, \nu, r_{})\ 
e^{\left[- {\textstyle \f{1}{2}} 
(\mu^2 - 2r_{} \mu \nu + \nu^2)/(1 - r_{}^2) \right]} \,.
\label{pdf10}
\end{eqnarray}
Here, $H_{p'q'}$ are the bivariate Hermite polynomials,
\ba
&&\hspace{-.7cm}
H_{p'q'}(\mu, \nu, r_{}) \equiv \nonumber\\
&&\hspace{-.7cm}
(-1)^{p'+q'} 
\f{\p^{p'}}{\p \mu^{p'}} \f{\p^{q'}}{\p \nu^{q'}} 
\exp{\left[- \f{\mu^2 - 2r_{} \mu \nu + \nu^2}{2(1 - r_{}^2)} \right]} 
\,.
\label{pdf11}
\ea 
We will demonstrate below that unfortunately this approach is not strictly
applicable for our case, although it could remain a very useful
approximation. 
The cumulant $\lam_{11}$ is the correlation coefficient of the fields
in question. Let us write $r_{} = \lam_{11}$. We have
\be 
r_{} = 1 - {\textstyle \f{1}{2}} \alpha \s^2 + \calO(\s^4) \,,
\label{pdf12}
\ee 
where $\alpha$ is defined by equation~(\ref{alpha1}). The correlation
coefficient is at the leading order equal to unity, what means that
$\de$ and $\te$ are strongly correlated in the weakly nonlinear
regime. For the time being, let us concentrate on the up-to-linear
terms in the expansion of the exponent in integral~(\ref{pdf8}). We
then have
\begin{eqnarray}
&&\hspace{-.7cm}
P(\mu,\nu) = \f{1}{2\pi (1 - r_{}^2)^{1/2} } \times\nonumber\\
&&\hspace{-.7cm}\left[ 
1 + \sum_{n = 3}^{\infty} (-1)^n \sum_{k = 0}^{n} 
\f{\lam_{(n-k)k}}{(n-k)! k!} 
\left(\f{\p}{\p \mu}\right)^{n-k} \left(\f{\p}{\p \nu}\right)^{k}
\right] \times \nonumber \\
&&\hspace{-.7cm}
\exp{\left[- \f{\mu^2 - 2r_{} \mu \nu + \nu^2}{2(1 - r_{}^2)} \right]}
+ \hbox{\rm other terms} 
\,.
\label{pdf13}
\end{eqnarray}
Including only the first term of the expansion yields the bivariate
Gaussian distribution, while including also the terms with $n=3$ in
the double series yields the second-order bivariate Edgeworth
expansion (Longuet-Higgins~1963). In our case $r_{}$ is given by
equation~(\ref{pdf12}), so we can write
\ba
&&\hspace{-.7cm}
f(\mu, \nu, r_{}) \equiv 
\exp{\left[- \f{\mu^2 - 2r_{} \mu \nu + \nu^2}{2(1 - r_{}^2)} \right]}
\nonumber\\
&&\hspace{-.7cm}= \exp{\left[- (x^2 + \nu^2)/2 \right]} + \calO(\s) \,,
\label{pdf14}
\ea
where
\be
x \equiv \f{\mu - \nu}{\alpha^{1/2} \s} \,.
\label{pdf15}
\ee 
Thus, at the leading order the dependence of $f$ on $x$ and $\nu$
factors out. This suggests a useful change of variables, $(\mu, \nu)
\to (x, \nu)$. From the chain rule we have $(\p/\p \mu)_{\vert_{\nu}} =
(\alpha^{1/2} \s)^{-1} (\p/\p x)_{\vert_{\nu}}$ and $(\p/\p \nu)_{\vert_{\mu}} =
(\p/\p
\nu)_{\vert_{x}} - (\alpha^{1/2} \s)^{-1} (\p/\p x)_{\vert_{\nu}}$. After some
algebra, this yields
\begin{eqnarray}
&&\hspace{-.7cm}
P(\mu,\nu) = \f{1}{2\pi \alpha^{1/2} \s} \times\nonumber\\
&&\hspace{-.7cm}\left[ 
1 + \sum_{n \ge 3,k\le n}
\f{(-1)^n \lan (\mu - \nu)^{n-k} \nu^k \ran_c}{(n-k)! k! \, 
\alpha^{(n-k)/2} \s^{n-k} } 
\left(\f{\p}{\p x}\right)^{n-k}\!\left(\f{\p}{\p \nu}\right)^{k}
\right] f
\nonumber \\
&&\hspace{-.7cm} + \; \hbox{\rm other terms} 
\,.
\label{pdf16}
\end{eqnarray}
Using equations~(\ref{pdf6}) and~(\ref{pdf14}), we can cast the last 
expression to the form
\begin{eqnarray}
&&\hspace{-.7cm}
P(\mu,\nu) 
= \f{1}{2\pi \alpha^{1/2} \s} \times\nonumber\\
&&\hspace{-.7cm}
\left[ 1 + \sum_{n \ge 3,k\le n}
\f{c_{nk}}{(n-k)! k! \, \alpha^{(n-k)/2}} 
H_{n-k}(x) H_k(\nu) \right] f  
\nonumber \\
&&\hspace{-.7cm} + \; \hbox{\rm other terms} 
\,,
\label{pdf17}
\end{eqnarray}
where
\be
c_{nk} \equiv \f{\lan (\de - \te)^{n-k} \te^k \ran_c}{\s^{2n-k}} 
\label{pdf18}
\ee
and $H_n$'s are the ordinary Hermite polynomials. The tight
correlation between $\de$ and $\te$ causes the $x$ and $\nu$
dependence of each corrective term to factor. This simplification,
however, comes at a price: we cannot formally truncate the series in
$n$. The leading-order contribution to the numerator of the
coefficient $c_{n0}$ is of the order of $\s^{2n}$, hence $c_{n0}$ is
of the order of unity, for all $n$. The numerator of $c_{n1}$ is
similarly of the order of $\s^{2n}$, hence $c_{n1}$ is of the order of
$\s$. For $k \ge 2$, the numerator of $c_{nk}$ is of the order of
$\s^{2n - 2}$, hence $c_{nk} \propto \s^{k - 2}$. Therefore, to
include all corrective terms up to, say, quadratic in $\s$, we can
truncate the series in equation~(\ref{pdf17}) at $k = 4$, but we
cannot truncate it at any $n$. 
Why is it so? The variables $\de$ and $\te$ are equal at first order,
so the variable $x \propto \de - \te$ is fully nonlinear. Thus, $\nu$
is a weakly non-Gaussian variable, but $x$ is a fully non-Gaussian
one. As a result, while the cumulants of $\nu$ form a hierarchy in
$\s$ the cumulants of $x$ do not and we need an infinite number of
terms in the Edgeworth expansion to describe the distribution of
$x$. Since this distribution can be derived from the joint $P(\de,
\te)$, the joint PDF cannot be formally Edgeworth-expanded (to finite
order) either.

\end{document}